\documentclass{ieeeaccess}
\pdfoutput=1
\usepackage{cite}
\usepackage{amsmath,amssymb,amsfonts}
\usepackage{algorithmic}
\usepackage{graphicx}
\usepackage{textcomp}

\usepackage[hyphens]{url}
\usepackage{graphicx}
\usepackage{csquotes} 
\usepackage{ragged2e}
\usepackage{amsthm}
\usepackage[per-mode = symbol]{siunitx}
\DeclareSIUnit\year{year}
\DeclareSIUnit\fatality{fatality}
\DeclareSIUnit\fatalities{fatalities}
\usepackage{verbatim}
\usepackage{array}
\usepackage{hyperref}

\usepackage{enumerate}
\usepackage[shortlabels]{enumitem}
\usepackage{microtype} 
\usepackage{adjustbox} 
\usepackage[all]{nowidow}

\usepackage{hologo}
\usepackage{listings}
\newcounter{requirement}[section]
\newenvironment{requirement}[1][]{\refstepcounter{requirement}\par\medskip
  \noindent \textbf{Requirement~\therequirement. #1} \rmfamily}{\medskip}

\def\BibTeX{{\rm B\kern-.05em{\sc i\kern-.025em b}\kern-.08em
    T\kern-.1667em\lower.7ex\hbox{E}\kern-.125emX}}
\begin{document}

\thispagestyle{empty}
\pagestyle{empty}
\twocolumn[
\begin{@twocolumnfalse}
	%
	%
	
	\Huge {IEEE copyright notice} \\ \\ 
	\large {\copyright\ 2024 IEEE. Personal use of this material is permitted. Permission from IEEE must be obtained for all other uses, in any current or future media, including reprinting/republishing this material for advertising or promotional purposes, creating new collective works, for resale or redistribution to servers or lists, or reuse of any copyrighted component of this work in other works.} \\ \\
	
	{\Large Published in \emph{IEEE Access}} \\ \\
	
	{\Large DOI: \href{https://doi.org/10.1109/ACCESS.2024.3372860}{10.1109/ACCESS.2024.3372860}} \\ \\
	
	Cite as:
	\vspace{0.1cm}
	
	\noindent\fbox{%
		\begin{minipage}{0.98\textwidth}%
			N.~Salem, T.~Kirschbaum, M.~Nolte, C.~Laltisch-Schneider, R.~Graubohm, J.~Reich, and M.~Maurer, ``{Risk} {Management} {Core} -- {Towards} an {Explicit} {Representation} of {Risk} in {Automated} {Driving},'' \emph{IEEE Access}, vol. 12, pp. 33200-33217, 2024. DOI: 10.1109/ACCESS.2024.3372860.
		\end{minipage}
	}
	\vspace{2cm}
	
\end{@twocolumnfalse}
]

\noindent%
\hologo{BibTeX}:

\noindent
	\begin{centering}
	\footnotesize
	\begin{lstlisting}[frame=single,linewidth=\textwidth]
            @article{salem_risk_2024,
	title = {Risk {Management} {Core}---{Toward} an {Explicit} {Representation} of {Risk} in {Automated} {Driving}},
	volume = {12},
	journal = {IEEE Access},
	author = {Salem, Nayel Fabian and Kirschbaum, Thomas and Nolte, Marcus and Lalitsch-Schneider, Christian and Graubohm, Robert and Reich, Jan and Maurer, Markus},
	year = {2024},
	pages = {33200--33217},
}
	\end{lstlisting}
\end{centering}


\history{Date of publication xxxx 00, 0000, date of current version xxxx 00, 0000.}
\doi{10.1109/ACCESS.2017.DOI}

\title{Risk Management Core -- Towards an Explicit Representation of Risk in Automated Driving}
\author{
\uppercase{Nayel Fabian Salem}\authorrefmark{1},
\uppercase{Thomas Kirschbaum}\authorrefmark{2},
\uppercase{Marcus Nolte}\authorrefmark{1},
\uppercase{Christian Lalitsch-Schneider}\authorrefmark{3},
\uppercase{Robert Graubohm}\authorrefmark{1},
\uppercase{Jan Reich}\authorrefmark{4},
and \uppercase{Markus Maurer}\authorrefmark{1}
}

\address[1]{Institute of Control Engineering, Technische Universit\"at Braunschweig, 38106 Braunschweig, Germany (e-mail: \{salem, nolte, graubohm, maurer\}@ifr.ing.tu-bs.de)}
\address[2]{Robert Bosch GmbH, 74003 Heilbronn, Germany (e-mail: thomas.kirschbaum@de.bosch.com)}
\address[3]{ZF Friedrichshafen AG, 88046 Friedrichshafen, Germany (e-mail: christian.lalitsch-schneider@zf.com)}
\address[4]{Fraunhofer Institute for Experimental Software Engineering IESE, 67663 Kaiserslautern, Germany (e-mail: jan.reich@iese.fraunhofer.de)}
\tfootnote{This work was supported by the German Federal Ministry for Economic Affairs and Climate Action within the PEGASUS Family project ``Verifikations- und Validierungsmethoden automatisierter Fahrzeuge im urbanen Umfeld (VVMethods)'',
a Successor Project to the project ``Projekt zur Etablierung von generell akzeptierten Gütekriterien, Werkzeugen und Methoden sowie
Szenarien und Situationen zur Freigabe hochautomatisierter Fahrfunktionen (PEGASUS)'' and ``Automatisierter Transport zwischen Logistikzentren auf Schnellstraßen im Level 4 (ATLAS-L4)''.
}

\markboth
{Salem \headeretal: Risk Management Core -- Towards an Explicit Representation of Risk in Automated Driving}
{Salem \headeretal: Risk Management Core -- Towards an Explicit Representation of Risk in Automated Driving}

\corresp{Corresponding author: Nayel Fabian Salem (e-mail: salem@ifr.ing.tu-bs.de).}

\begin{abstract}
While current automotive safety standards provide implicit guidance on how unreasonable risk can be avoided, manufacturers are required to specify risk acceptance criteria for Automated Driving Systems (SAE~Level~3 and higher). However, the ‘unreasonable’ level of risk of Automated Driving Systems is not yet concisely defined. Solely applying current safety standards to such novel systems could potentially not be sufficient for their acceptance. As risk is managed with implicit knowledge about safety measures in existing automotive standards, an explicit alignment with risk acceptance criteria is challenging. Hence, we propose an approach for an explicit representation and management of risk, which we call the Risk Management Core. The proposal of this process framework is based on requirements elicited from current safety standards and is applied to the task of specifying safe behavior for an Automated Driving System in an example scenario.
\end{abstract}

\begin{keywords}
Automated Driving, Behavior Specification, Risk, Risk Management, Safety
\end{keywords}

\titlepgskip=-15pt

\maketitle

\section{Introduction}
\label{sec:introduction}
\PARstart{A}{ccording} to the Commission Implementing Regulation (EU) 2022/1426~\cite{noauthor_commission_2022} the \enquote{[...] manufacturer [of an Automated Driving System (ADS)] shall define the acceptance criteria from which the validation targets of the ADS are derived to evaluate the residual risk for the ODD [(i.\,e. the Operational Design Domain)] [...]}. However, current automotive safety standards such as ISO~26262~\cite{noauthor_road_2018} and ISO~21448~\cite{noauthor_road_2022} only provide implicit guidance on how risk acceptance criteria can be met in a safety life cycle of road vehicles. These domain-specific standards apply guidelines from their underlying safety standard IEC~61508~\cite{noauthor_functional_2010} to the automotive domain. Whereas IEC~61508 describes risk management activities that explicitly reduce the residual risk of a system of interest to a reasonable level, ISO~26262 specifies a number of measures that are implicitly known to adequately reduce the risk of a system (or \emph{item}).

For conventional vehicles and vehicles with driver assistance systems (SAE~Level~2 and lower) applying best practices~\cite{knapp_code_2009, cao_l3pilot_2021} led to general acceptance as the reasonable level of risk for these systems or at least adequate measures to achieve that level were implicitly known: As long as a driver assistance system can be overruled by the driving person or only engages in situations that would be uncontrollable for a human driver, the capabilities of a human driver to fulfill the dynamic driving task directly serve as a proxy reference for accepted risk. However, for Automated Driving Systems (SAE~Level~3 and higher) neither the reasonable level of risk is specified~\cite{junietz_microscopic_2019} nor is there a common understanding of adequate measures leading to such a level of risk.
Managing risk for automated vehicles poses multiple challenges, which are: 
\begin{itemize}
    \item [C1:] Defining valid risk acceptance criteria~\cite{junietz_macroscopic_2019},
    \item [C2:] analyzing the risk introduced by an automated vehicle~\cite{de_gelder_risk_2021}, 
    \item [C3:] specifying measures for aligning the actual risk introduced by an automated vehicle with the specified risk acceptance criteria~\cite{filip_derivation_2022},
    \item [C4:] validating the specified measures in an open traffic context (both pre- and post-deployment~\cite{galbas_safeguarding_2022}, and
    \item [C5:] communicate risk management with all relevant stakeholders (e.g. type approval agencies)~\cite{noauthor_cross-domain_2023}.
\end{itemize}

The analysis and definition of risk acceptance criteria~(C1), which involves negotiations on a societal level, is outside the scope of this work. Instead, we propose a framework that contributes to an explicit management of risk by supporting the analysis of risk introduced by an automated vehicle~(C2) and the specification of risk reduction measures~(C3). We call this framework the \emph{Risk Management Core (RMC)} as it describes a general approach for managing risk and it can be applied to multiple activities during a safety life cycle. It is designed to align the risk that is introduced by an automated vehicle (both prospectively and retrospectively) with specified risk acceptance criteria. The validation and bi-directional communication of required and implemented risk reduction measures (C4 and C5) is a challenge we will address as part of future work.

This paper addresses challenges C2 and C3 by:
\begin{itemize}
    \item eliciting requirements for risk management frameworks from existing safety standards (\autoref{sec:req}),
    \item proposing an ontology that captures necessary terminology for risk management (\autoref{sec:term}),
    \item describing the process framework \emph{Risk Management Core} (\autoref{sec:rmc}), and
    \item applying the Risk Management Core to the novel task of specifying safe behavior of Automated Driving Systems (\autoref{sec:risk_bspec}).
\end{itemize}

As a result, this paper also contributes to the challenge of communicating the achieved level of safety of an ADS in public. Safety is a term, where there is no common understanding about its meaning~-- especially among different stakeholders~\cite{salem_safety_2023}. Automotive safety standards and reports relevant for automated vehicles such as ISO~26262~\cite{noauthor_road_2018}, ISO~21448~\cite{noauthor_road_2022}, and ISO/TR~4804~\cite{noauthor_road_2020} use a definition of safety that refers to the absence of unreasonable risk. While the definition and the respective guidelines clearly focus on the risk of physical injury~\cite{noauthor_road_2018}, some stakeholders might consider further factors, such as damage to property, in their determination of a reasonable level of risk~\cite{salem_safety_2023}. Hence, arguing safety in a way that convinces multiple stakeholders becomes challenging.
-- Particularly given the lack of common understanding of valid risk acceptance criteria that define a reasonable level of risk. Thus, in our view, an explicit risk representation supports the alignment of actual risks with various risk acceptance criteria that shall be specified for Automated Driving Systems~\cite{noauthor_commission_2022, noauthor_road_2022} in a socially acceptable manner. As a result, risk can be addressed and managed by manufacturers throughout the safety life cycle of an automated vehicle. Explicit documentation of risk management activities will not achieve the stakeholder-independent harmonization of the terms ‘safety’ and ‘risk’. However, it contributes to transparent communication between different stakeholders~\cite{slovic_perceived_1993}.

\section{Related Research} \label{sec:relwork}
The challenges with respect to conceptualizing and applying risk management are not exclusive to the field of automated driving. Fischhoff states that \enquote{every year (or, perhaps,
every day), some new industry or institution discovers that it, too, has a risk problem}~\cite{fischhoff_risk_1995}. According to Fishhoff, exposing conflicts is desirable in order to conduct effective risk management (and risk communication).

From an ethical perspective on automated driving Godall~\cite{goodall_away_2016} argues that, historically, there has been a focus on trolley problems in order to discuss reasonable levels of risk. While acknowledging the use of such synthetic edge cases, Godall points out that risk management provides beneficial properties like transparency and adjustability for the assessment of different behaviors. In contrast to the consideration of trolley problems, a risk-based approach always leads to the recommendation of some action instead of solely making conflicting rules evident in a certain context. Therefore, a more nuanced decision-making is possible which Godall argues to be comparable with the behavior of human drivers. We follow the line of argumentation of both, Fischhoff and Godall, which leads us to our proposal of explicitly representing risk in the context of automated driving.

Current automotive safety standards provide guidance on how to conduct risk management. However, following standards does not automatically guarantee that unreasonable risk is avoided. This is particularly true for systems that operate in complex environments such as Automated Driving Systems. Fowler~\cite{fowler_iec_2022} discusses the shortcomings of safety standards in the transport sector. He focuses on their compliance with their foundational standard, ISO/IEC~61508~\cite{noauthor_functional_2010}. In the context of risk management this means that these standards have focused more on the reliability rather than on risk-reduction properties of safety-related systems. As a result, we do not only consider ISO~26262~\cite{noauthor_road_2018} and ISO~21448~\cite{noauthor_road_2022} but at the same time show compliance of our proposal with foundational standards such as ISO/IEC~61508 and ISO~Guide~51~\cite{noauthor_safety_2014}.

Other research publications that we considered in our work, focus on distinct aspects in risk management and therefore do not provide a holistic perspective. For example, publications we found in the literature contribute to risk management by either 
\begin{itemize}
    \item discussing the determination of accepted levels of risk~\cite{junietz_macroscopic_2019, wachenfeld_how_2017, filip_derivation_2022, favaro_exploring_2021},
    \item proposing approaches for risk analysis~\cite{de_gelder_risk_2021},
    \item focusing on the verification and validation of a specified level of risk~\cite{kauffmann_positive_2022, fahrenkrog_implications_2023}, or
    \item proposing approaches for maintaining a certain level of risk at runtime~\cite{geisslinger_autonomous_2021, reich_sinadra_2020, nolte_supporting_2020, harrison_safety_2008}.
\end{itemize}

However, to our knowledge, there is no published work on a risk management framework for automated driving that focuses on an explicit representation of risk and which allows to argue for compliance with automotive and non-automotive safety standards.

\section{Requirements for Risk Management Frameworks in the Context of Automated Driving} \label{sec:req}
In the previous sections we pointed out the need for a framework that supports an explicit representation of risk for Automated Driving Systems. Hence, in this section we elicit requirements for such a framework from existing safety standards. These requirements are not exhaustive as we do not cover all international standards that address risk management. However, we take selected standards as a basis to propose a general risk management process framework.

In the automotive sector, ISO~26262 is an important reference point for other safety standards such as ISO~21448 and ANSI/UL~4600~\cite{noauthor_standard_2020}. ISO~26262 provides a framework for managing risk implicitly in order to achieve functional safety: First of all, ISO~26262 does not specify explicit risk acceptance criteria. At the same time, to allow the argumentation for a functionally safe system, it is necessary to perform a hazard analysis and risk assessment and afterwards reduce the identified potential risk to a reasonable amount. However, the contribution of the specified measures to the necessary risk reduction in order to achieve a "reasonable" level of risk is not quantified, either. The implicitness of the way risk is managed in ISO~26262 becomes even more evident when examining the parameters that are provided for the analysis of hazardous events and the definition of safety goals. Hazardous events shall be classified by using classes for the severity of potential harm~(S), the exposure to an operational situation~(E), and the controllability of a hazardous event~(C) by the driver or other persons involved. As a result of this classification, safety goals shall be defined and assigned with a respective automotive safety integrity level (ASIL). This ASIL depends on the result of the classification of the hazardous events that are addressed by the safety goal. While clearly specifying organizational and process requirements as well as hardware-related metrics, which regard the integrity level of safety goals and safety requirements, ISO~26262 does neither explicitly state the underlying risk acceptance criteria used for the definition of ASIL nor the necessary quantitative risk reduction that has to be achieved. Hence, it is left open to the readers of the standard and therefore implicit, how the implementation of the dedicated ASIL-dependent measures leads to the absence of an unreasonable level of risk.

Fowler~\cite{fowler_iec_2022} points out the inadequacy of safety standards from the transport sector (focusing on railway and air traffic management) concerning conformance with IEC~61508~\cite{noauthor_functional_2010}. He elaborates the distinction to be made between engineering practices that address the reliability of safety critical systems and those that affect their risk-reduction properties.

Since ISO~26262 is an important safety standard in the automotive domain, we analyze it with respect to risk managing activities for assuring functional safety. More specifically, we focus on Part~1 (Vocabulary) and Part~3 (Concept phase), as they define and describe risk management guidelines. Subsequently, we elicit requirements from IEC~61508 to address the analyzed shortcomings in its application to the automotive domain with respect to explicit risk management. 

The scope of ISO~26262 is defined in Part~1. It is stated that the document addresses possible hazards caused by malfunctioning behavior of safety-related electrical and/or electronic (E/E) systems. While the scope excludes certain causes of hazards (e.\,g. hazards caused by the intended behavior) it does not limit the general scope of how risk reduction is specified. Part~3, Clause~6 \emph{(Hazard analysis and risk assessment)} and Clause~7 \emph{(Functional safety concept)} provide guidelines regarding challenges~C2 (analyzing risk) and~C3 (specifying countermeasures).

Firstly, we take the objectives formulated in ISO~26262~\cite[Part~3,~6.1]{noauthor_road_2018} and translate them into requirements for a risk management framework in the context of automated driving.

\begin{requirement}
Hazardous events shall be identified. \label{req1}
\end{requirement}

ISO~26262-1 defines a hazardous event as the combination of a hazard and an operational situation. Given the scope of functional safety, we can understand that malfunctioning behavior~(i.\,e. a failure or unintended behavior of a system caused by hardware or software faults~\cite[Part~1,~3.88]{noauthor_road_2018}) can cause hazards~(i.\,e. potential sources of harm~\cite[Part~1,~3.75]{noauthor_road_2018}). Hazards on the other hand manifest within operational situations of the system of interest. Hence, the identification of deviations from the desired vehicle behavior \cite{graubohm_towards_2020} and emerging hazardous events (Requirement~\ref{req1}) build the foundation for reducing risk to a reasonable level in the scope of ISO~26262.

\begin{requirement}
The risk of hazardous events shall be assessed based on the probability of the occurrence of harm and the severity of that harm. \label{req2}
\end{requirement}

Requirement~\ref{req2} can be detailed by further analyzing ISO~26262. Since risk is defined as a \enquote{combination of the probability of occurrence of harm and the severity of that harm}~\cite[Part~1,~3.128]{noauthor_road_2018}, we elicit requirements \ref{req3}-\ref{req5}.

\begin{requirement}
The harm potentially resulting from a hazardous event shall be identified. \label{req3}
\end{requirement}

\begin{requirement}
The severity of harm potentially resulting from a hazardous event shall be estimated. \label{req4}
\end{requirement}

\begin{requirement}
The probability of occurrence of harm potentially resulting from a hazardous event shall be estimated.%
\label{req5}%
\end{requirement}

In ISO 26262, risk assessment (including the estimation of the probability of occurrence of harm and severity of that harm) is described implicitly: Hazardous events are classified by estimating the parameters \emph{Severity}, \emph{Exposure}, and \emph{Controllability}. Based on these parameters and the resulting classification, ISO~26262 requires the determination of an ASIL. ASILs specify necessary ISO~26262 requirements and safety measures to apply in order to avoid unreasonable risk. Here, we refer back to Fowler~\cite{fowler_iec_2022}, who points out the clear separation of safety integrity and risk reduction made in IEC~61508~\cite{noauthor_functional_2010}. In order to reduce the risk inherited by a system (i.\,e. \emph{equipment under control} (EUC)), measures shall be taken. These measures can be realized by safety functions of the respective system. Integrity is defined as \enquote{the probability of a [...] safety-related system satisfactorily performing the specified safety functions under all the stated conditions, within a stated period of time}~\cite{noauthor_functional_2010}. Therefore, the integrity contributes (negatively) to the risk reduction of specified safety functions. In contrast, ISO~26262 merges risk reduction and integrity into the process of determining an ASIL and later defining safety goals. While this approach was successfully implemented for existing driver assistance systems, we further decompose requirements specified in ISO~26262 into requirements~(\ref{req8}-\ref{req10}) that we identify from IEC~61508 in order to account for the challenge of introducing a novel technology such as Automated Driving Systems.

\begin{requirement}
Safety measures shall be specified that lead to an acceptable level of risk~(i.\,e. a required level of safety). \label{req8}
\end{requirement}

\begin{requirement}
The risk reduction achieved by a specified safety measure shall be assessed. \label{req9}
\end{requirement}

\begin{requirement}
The probability of satisfactorily performing specified safety relevant functionality (i.\,e. its safety integrity) under all the stated conditions within a stated period of time shall be assessed. \label{req10}
\end{requirement}

Even though IEC~61508 Part~5 Annex~A is informative and thus does not contain normative requirements, the necessity of defining risk acceptance criteria is stated (Requirement~\ref{req11}). For the automotive domain, this is also specified in ISO~21448~\cite[Clause 6.1]{noauthor_road_2022} (normative).

\begin{requirement}
The required level of safety (i.\,e. the reasonable level of risk) shall be specified by defining risk acceptance criteria for the system of interest, the people, property, or the environment facing the risk, respecting valid societal moral concepts within the target market. \label{req11}
\end{requirement}

These general requirements, which are based on IEC~61508, cover most of the objectives mentioned in ISO~26262~\cite[Part~3,~7]{noauthor_road_2018}. One specific concept, which is added in ISO~26262, is the concept of \emph{safety goals} (Requirement~\ref{req12}). They are defined as top-level safety requirements and capture the targeted risk reduction to a reasonable level including the necessary integrity level. Note that while ISO~26262 focuses exclusively on functional safety, we adopt the concept of safety goals more generally.

\begin{requirement}
Safety goals with their corresponding integrity requirements shall be formulated related to the prevention or mitigation of the hazardous events in order to avoid unreasonable risk. \label{req12}
\end{requirement}

Our analysis of ISO~21448 does not lead to additional requirements regarding the representation of risk. In contrast to functional safety, SOTIF (i.\,e. safety of the intended functionality) requires the consideration of specification and performance insufficiencies. For automated driving, this is an essential extension of the scope for which risk management shall be applied. However in terms of risk evaluation, ISO~21448~\cite[Clause 6.4]{noauthor_road_2022} mostly refers to ISO~26262.

ANSI/UL~4600 \cite[Clause 6]{noauthor_standard_2020} covers risk assessment explicitly. Here, different terminology is used to express similar concepts to those presented in IEC~61508. \enquote{Calculated probability of incident occurrence yields both probability and confidence for entire causal chain.}~\cite[Clause~6.1.1.3~b]{noauthor_standard_2020} Safety functions requirements and safety integrity requirements from IEC~61508 address both aspects: probability and confidence for the entire causal chain. Previous requirements we elicited mostly considered hazardous behavior that leads to hazardous events. Additionally, ANSI/UL~4600 (as well as ISO~26262) requires the identification and documentation of hazards (Requirement~\ref{req13}). The documentation of hazards in a hazard log (Requirement~\ref{req14}) also enables life cycle monitoring and risk management as required by ISO/IEC/IEEE~16085~\cite[Clause~6]{noauthor_systems_2021}.

\begin{requirement}\RaggedRight
\enquote{Potentially relevant hazards shall be identified.}~\cite[Clause 6.3.1]{noauthor_standard_2020} \label{req13}
\end{requirement}

\begin{requirement}
A hazard log shall be created, which lists identified hazards and their mitigation status. \label{req14}
\end{requirement}

Another normative source of requirements for risk management frameworks we found to be highly relevant is the ISO/IEC~Guide~51~\cite{noauthor_safety_2014}. The standard provides guidelines especially for drafting standards that cover safety-related aspects. Most of the requirements we have already listed so far can also be elicited based on ISO~Guide~51. An additional requirement that was implicitly covered by the requirements so far addresses the risk management activity of risk evaluation (Requirement~\ref{req15}).

\begin{requirement}
Risk evaluation shall be conducted, which determines the necessary risk reduction from the initial level of risk associated with the system of interest to the acceptable level of risk. \label{req15}
\end{requirement}

\Figure[ht](topskip=0pt, botskip=0pt, midskip=0pt)[width=1.6\columnwidth]{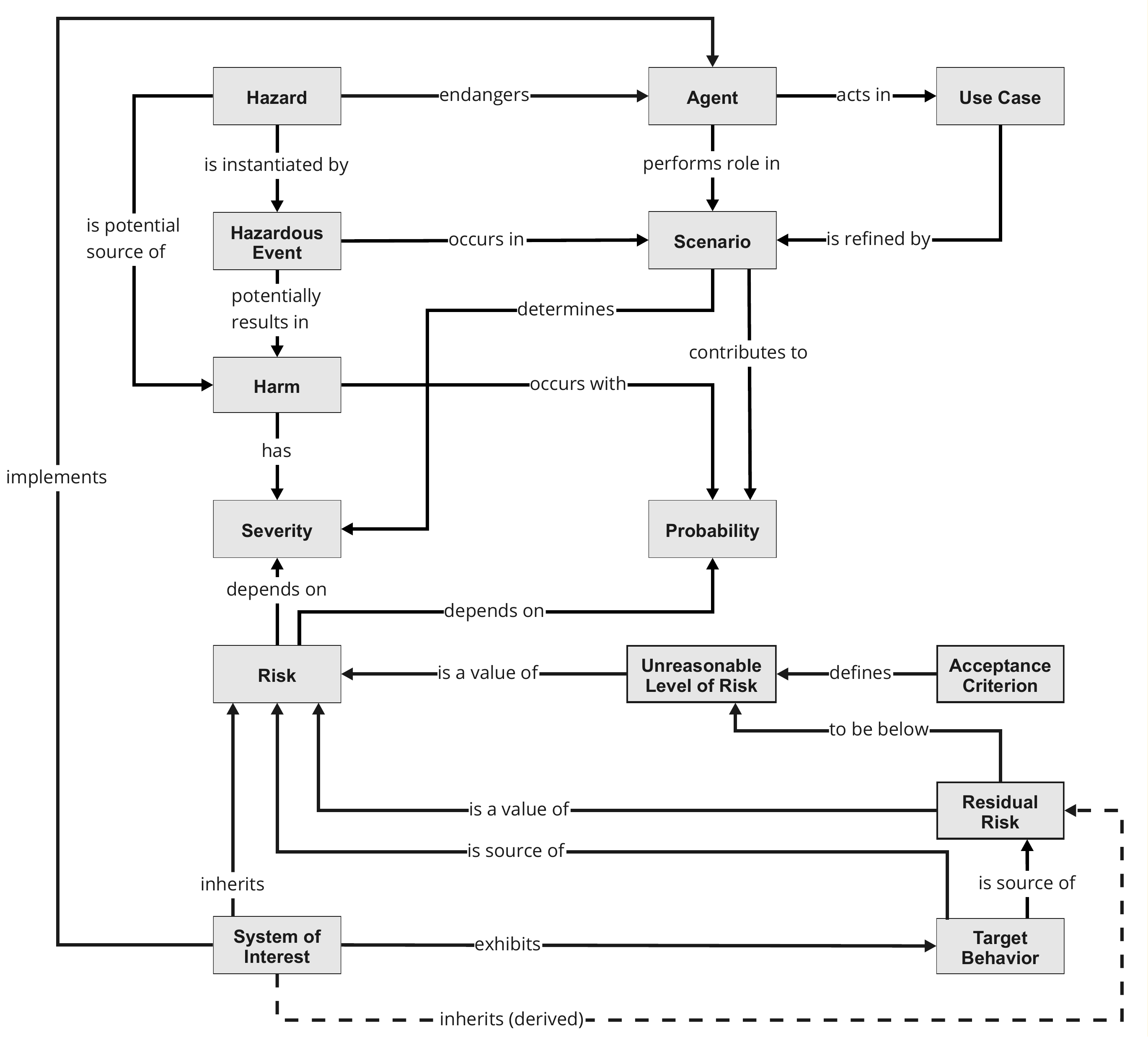}
{This risk assessment ontology provides concepts and their relations for describing activities related to risk analysis and risk evaluation. Based on ISO~Guide~51, risk analysis entails the identification of hazards and the estimation of risk. Risk evaluation is concerned with comparing the assessed residual risk inherited by a system of interest with an acceptable level of risk.\label{fig:risk_assess}}

\section{Terminology} \label{sec:term}
Based on the derived requirements for risk management frameworks in the context of automated driving, we designed an ontology that aims to merge the terminology used in the standards we have analyzed so far. As the ontology provides a means to describe concepts in the context of risk management it contributes to an explicit representation of risk in automated driving. 

Following the separation of activities described in ISO~31000, we will lay out necessary terminology in two steps: 
\begin{enumerate}
    \item risk assessment (\autoref{fig:risk_assess})
    \item risk treatment (\autoref{fig:risk_treat}).
\end{enumerate}

\subsection{Risk Assessment Ontology}
Part of our motivation for proposing the Risk Management Core is to support an argumentation for the absence of unreasonable risk emerging from an Automated Driving System. This requires an explicit alignment of actual risk with accepted risk. Hence, we first refer to the concept of \emph{risk} in \autoref{fig:risk_assess}. Risk is defined as the \enquote{combination of the probability of harm and the severity of that harm} in ISO~Guide~51. Thus, risk \emph{depends on} the \emph{severity} and \emph{probability} of \emph{harm}. 
With respect to the concept of harm, existing (automotive) safety standards refer to physical injury. However, stakeholders may interpret relevant harm differently. In our view, a concretization of the concept of harm is not required in the generic concept of the Risk Management Core but yields the possibility to address different stakeholder interpretations of the concept of risk.

Both, the probability of occurrence and the severity of harm for quantifying risk are determined by the specific context in which the harm potentially occurs. Following the definition of Ulbrich~\emph{et~al.}~\cite{ulbrich_defining_2015}, which is adopted by ISO~21448, we refer to the temporal development of a context as a \emph{scenario}. The hazardous event model introduced in ISO~21448 further defines that an event that potentially leads to harm in a scenario is called a \emph{hazardous event}. As a result, the number of all known (hazardous) scenarios can be used for an estimation of the risk inherited by the system of interest.

For a hazardous event to be identified as such, it is necessary to be aware of \emph{potential sources of} harm. These sources are defined as \emph{hazards}~\cite{noauthor_road_2018, noauthor_functional_2010, noauthor_safety_2014, noauthor_road_2022}. Hence, a hazardous event \emph{is an instance of} a hazard in a scenario. It is characterized by people, property, or the environment facing one or more hazards~\cite{noauthor_safety_2014, noauthor_functional_2010}. This terminology deviates from the use in the automotive context of functional safety in so far that ISO~26262 defines a hazardous event to occur in an operational situation. We assume this difference to be negligible, since the use of the concepts \enquote{operational situation} and \enquote{scenario} are similar in the context of hazard analysis and risk assessment activities (cf. \cite[Part~1,~3.104]{noauthor_road_2018}).

\Figure[ht](topskip=0pt, botskip=0pt, midskip=0pt)[width=1.7\columnwidth]{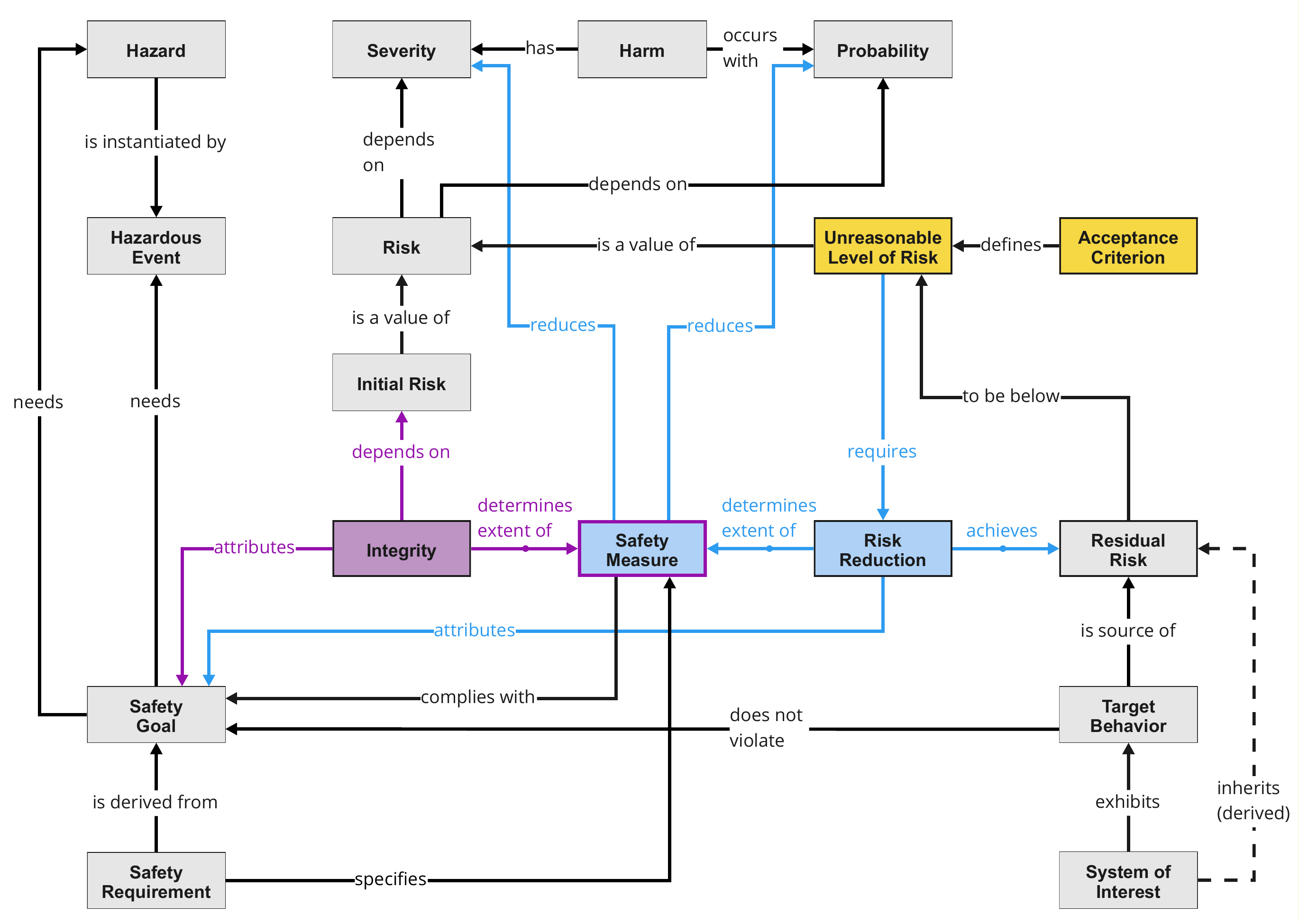}
{Risk treatment as defined in ISO~31000 succeeds the activity of risk assessment. \enquote{The purpose of risk treatment is to select and implement options for addressing risk.}~\cite{noauthor_risk_2018} In the context of automated driving, we include terms from ISO~26262 such as safety goals and requirements. Yellow color indicates concepts related to risk acceptance criteria. Blue color indicates concepts related to risk reduction. Purple color indicates safety-integrity-related concepts.\label{fig:risk_treat}}

As risk can more generally be expressed as an effect of uncertainty on objectives~\cite{noauthor_risk_2018} and as an objective is expressed by a subject, we include the concept of an \emph{agent} in the ontology, who is \emph{endangered} by a potential source of harm~(i.\,e. a hazard). In contrast to ISO~31000, which mostly refers to an organization's objectives being subject to uncertainty, we adopt the scope of ISO~26262 and ISO~21448 for the domain of automated driving. While an organization can also be referred to as an agent, we limit the interpretation to an individual subject that is a part of public road traffic for the scope of this paper. ISO~26262 \cite[Part 3, Clause 6.4.2.3]{noauthor_road_2018} states that hazards shall be identified at the vehicle level. Hence, the proposed ontology specifies that a system of interest implements an agent (i.\,e. a vehicle moving in traffic), which \emph{acts in} a use case. It is also the subject of the risk management activities as it \emph{exhibits} the specified \emph{target behavior}. The respective system boundary is therefore not necessarily specified by considering an agent but could, for example, cover a valet parking system including an ADS-equipped vehicle~\cite{noauthor_taxonomy_2021} supported by systems in the car park. This terminology also accounts for risk management conducted in the scope of subsystems of an ADS-EV such as a perception system. In order to account for the difference between a hazard as a potential source of harm and a hazardous event occurring in a specific scenario, we differentiate \emph{use cases} and scenarios. A use case \emph{subsumes} a suite of scenarios as it is defined in ISO~21448. Hence, an agent \emph{performs a role} in a scenario while acting in a use case.

Further terminology could be added based on, for example, the Rapid Exchange of Information System (RAPEX)~\cite{noauthor_rapex_nodate}, for which guidelines were set out by the European Commission in its Decision 2010/15/EU of December 16, 2009 and superseded by the Commission Implementing Decision (EU) 2019/417 of November 8, 2018~\cite{noauthor_commission_2019}. RAPEX suggests the use of \emph{danger} as a concept supporting risk assessment. However, no clear definition of danger is given. Based on our analysis, it thus seems to be redundant to the concepts of hazard, harm, and hazardous event. Thus, we do not adopt danger as a concept in our ontology, although the term may be useful in other applications. In a special report for the German parliament the difficulty of separating concepts such as danger and hazard across multiple domains such as the legal and the engineering community is underlined~\cite[in German]{rehbinder_sondergutachten_1999}.

All concepts we have elaborated on so far, can be subordinated to the activity of risk analysis, which is defined as a \enquote{systematic use of available information to identify hazards and to estimate the risk}~\cite{noauthor_risk_2018} in ISO~31000. Furthermore, we include concepts regarding risk evaluation in our ontology. Risk evaluation is a \enquote{procedure based on the risk analysis to determine whether tolerable risk has been exceeded}~\cite{noauthor_risk_2018}.

Based on Requirement~\ref{req11}, elicited from IEC~61508 and ISO~21448, a risk management framework in the context of automated driving requires including \emph{risk acceptance criteria}. Such criteria \emph{define} the absence of an \emph{unreasonable level of risk}, which \emph{is a value of} risk. ISO~Guide~51 names such a level of risk tolerable instead of acceptable. Hence, acceptable, tolerable, and reasonable are used synonymously in our terminology for risk management. Challenges with respect to the definition of unreasonable risk based on acceptance criteria such as the positive risk balance are elaborated by Favar\`o~\cite{favaro_exploring_2021}.

ISO~Guide~51 acknowledges that \enquote{some level of risk is inherent in products or systems.}~\cite{noauthor_safety_2014} This is especially relevant for Automated Driving Systems (SAE~Level~3 and higher) as they act in an open traffic context~\cite{nolte_representing_2018}. Hence, a system of interest \emph{inherits} risk even if it \emph{exhibits} a specified \emph{target behavior}. This behavior, which is specified for the agent performing a role in a scenario, \emph{is a} (i.\,e. one possible source) \emph{source of} some level of \emph{residual risk}. The residual risk \emph{is a value of} risk and \emph{shall be below} the acceptable level of risk in order for the system of interest to be deemed safe.

\subsection{Risk Treatment Ontology}
Reducing the difference between the \emph{residual risk} that is \emph{inherited by} the system of interest and the respective acceptable level of risk shall be \emph{achieved} by a specified amount of \emph{risk reduction}~(cf.~\autoref{fig:risk_treat}). The \emph{required} risk reduction \emph{determines the extent of} adopted \emph{safety measures}, which \emph{reduce} the \emph{severity} and/or the \emph{probability} of harm in order to satisfy the risk acceptance criteria. Based on Requirement~\ref{req10}, the extent of the safety measures is also \emph{determined by} a certain \emph{integrity}. Whereas the necessary risk reduction depends on the unreasonable level of risk, the integrity \emph{depends on} the \emph{initial risk} inherited by the system of interest without adopted safety measures. ISO/IEC~61508 defines integrity (cf. \autoref{fig:fowl}) as the \enquote{probability of an E/E/PE safety-related system satisfactorily performing the specified safety functions under all the stated conditions within a stated period of time.} \cite{noauthor_functional_2010}

In line with Requirement~\ref{req8}, \emph{safety measures} are specified that are necessary to fulfill the risk acceptance criteria. The explicit specification of measures and their contribution to the overall risk reduction is an essential step in explicitly representing risk. In contrast, ISO~26262 specifies measures as a result of an ASIL-classification that -- based on implicit knowledge -- lead to a reasonable level of risk. Furthermore, we adopt the term \emph{safety goal} as it is used in ISO~26262 in so far that a safety goal \emph{needs} hazards and hazardous events. Safety goals are thereby top-level safety requirements and are \emph{attributed with} both the necessary risk reduction and integrity. Finally, as the target behavior of the system of interest is being a source of residual risk, it \emph{shall not violate} the safety goals.

\section{Risk Management Core} \label{sec:rmc}
After having laid out the fundamental concepts and their relations that, in our view, provide a solid basis for explicit risk management, we propose the Risk Management Core (RMC) as a process framework for managing risk in automated driving. The framework aims at aligning actual risk and accepted risk in an iterative fashion. As risk management is a task being performed on different system levels~(e.\,g., operational, functional, technical) and throughout the life cycle of products and organizations we do not claim here that the proposed process framework suits each of the respective needs. However, based on normative requirements elicited in \autoref{sec:req} we propose the Risk Management Core as a framework that enables risk management in the context of automated driving. In this section we will present the framework with its essential activities and artifacts (\autoref{fig:rmc}). In the following section we will apply the Risk Management Core to the challenge of specifying safe behavior of an automated vehicle.

\Figure[ht](topskip=0pt, botskip=0pt, midskip=0pt)[width=2\columnwidth]{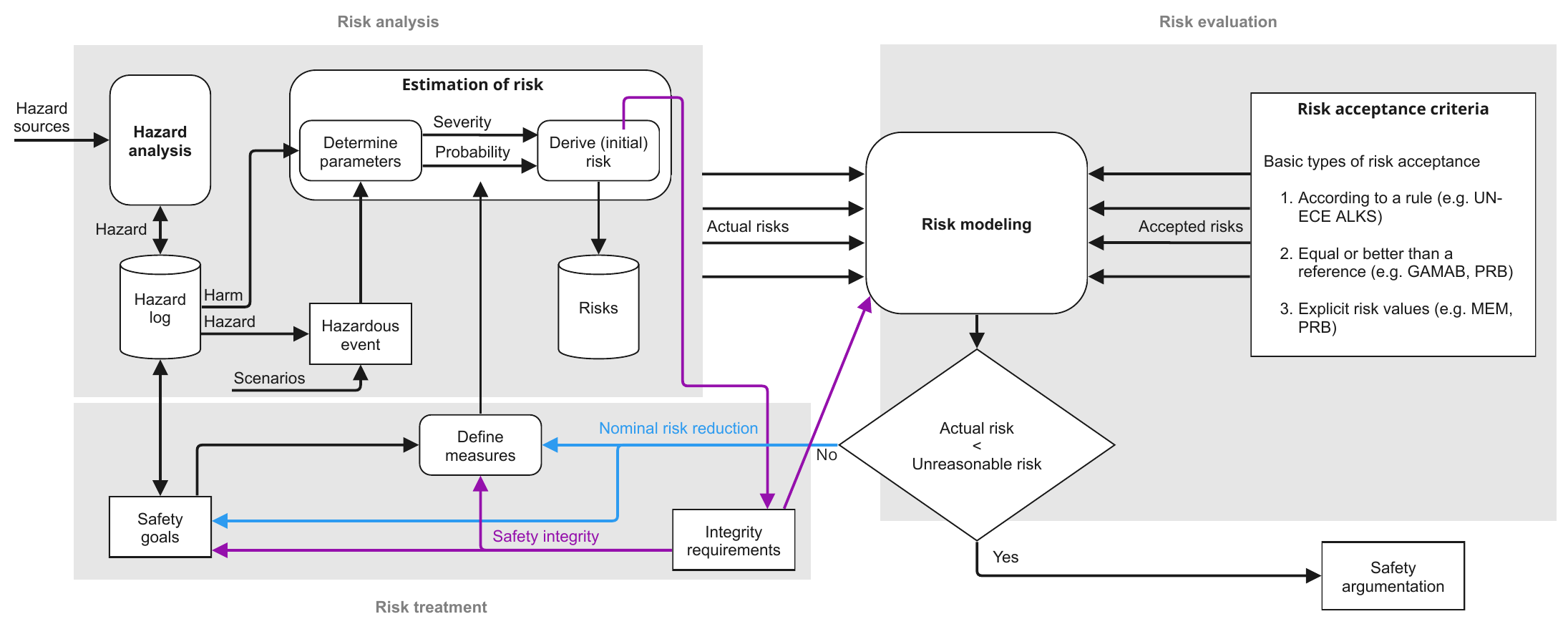}
{The Risk Management Core is an iterative process framework to align actual risk with accepted risk. This process framework is composed by risk analysis, risk evaluation and risk treatment.\label{fig:rmc}}

A prerequisite for using the Risk Management Core is basic knowledge about \emph{hazard sources}, which can be acquired, for example, by studying accident databases or based on expert analyses (cf. Neurohr~\emph{et~al.}~\cite{neurohr_criticality_2021}). Hazard sources are necessary to identify potential sources of harm~(i.\,e. hazards). Furthermore, in order to enable the alignment of actual and accepted risk, \emph{risk acceptance criteria} have to be given, which are valid for the determined desired operational environment of the Automated Driving System. Such acceptance criteria represent the absence of an unreasonable level of risk \cite{noauthor_road_2022}. Finally, a \emph{scenario catalog} for which the actual risk of the system shall be managed has to be available. 

Risk emerging from insufficiencies regarding these inputs is not in the scope of this paper. Insufficiencies could, for example, regard the incompleteness of the scenario catalog or incomplete knowledge about hazard sources. Note however, that the systematic management and explicit representation of risk facilitates the identification of both specification and performance insufficiencies. Furthermore, life cycle processes can be supported by integrating them with risk management activities by taking preventive measures and acting on in-field information \cite{noauthor_systems_2021}.

Generally, the Risk Management Core consists of three elements. Each of these elements are in line with the process described in ISO~Guide~51. Firstly, \emph{risk analysis} contains the steps of \emph{hazard analysis} and \emph{risk estimation}. Secondly, \emph{risk evaluation} is performed by comparing \emph{actual risk} and \emph{accepted risk}. The activity of risk assessment subsumes risk analysis and risk evaluation. Thirdly, \emph{risk treatment} shall address the difference identified in the risk evaluation step. ISO~31000 also defines further activities that are necessary to conduct risk management in a socio-technical context. \emph{Communication~\&~Consulting} as well as \emph{Monitoring~\&~Review} are specified as exclusive tasks. These tasks become increasingly important when considering life cycle risk management~\cite{noauthor_systems_2015, noauthor_systems_2021}. However, in the scope of this paper we focus on risk analysis, risk evaluation and risk treatment.

\subsection{Risk analysis}
As the overall aim of the Risk Management Core is the iterative alignment of actual risk and accepted risk, the process begins with analyzing risk emerging from the system of interest. In contrast to ISO~31000, we define risk identification to be a part of risk analysis. ISO~Guide~51 supports this way of structuring the risk management process. Hence, the step of risk analysis begins with the consideration of hazard sources. As defined in \autoref{sec:term}, a hazard is a potential source of harm that endangers some agent in the context of a use case. Previously recorded accidents and expert knowledge about the emergence of accidents can thus indicate causality between context conditions and harm~(cf.~\cite{ericson_hazard_2005}). Analyzing such causalities between cause and effect does neither require detailed technical knowledge about the agent nor about its operational environment. As a result, a \emph{hazard log} can be filled with an initial set of \emph{hazards} and their corresponding \emph{harms}.

Scenario-based approaches for safety assurance are intended to structure the operational environment of Automated Driving Systems. Thereby, they provide means to identify hazardous events and assess their respective risk. This knowledge can support the argumentation of the absence of unreasonable risk~\cite{noauthor_road_2021}. However, due to the necessary assumptions in the specification of a scenario catalog (especially when specifying logical or concrete scenarios) there is risk emerging from specification insufficiencies. In the Risk Management Core, known scenarios are used to identify \emph{hazardous events} as a combination of specific circumstances in a scenario and a known hazard. If an unsafe unknown scenario is encountered during operation, respective monitors \cite{mauritz_engineering_2020} can be utilized in future iterations of risk management.

According to UL~4600~\cite{noauthor_standard_2020}, identified hazardous events and known kinds of harm shall be logged as part of a hazard log. Such a log supports multiple aspects of risk management. As risk management is an iterative task, a record of the already acceptably addressed hazards and those that require further measures to be taken contributes to an explicit documentation of the risk management status. Furthermore, risk management is conducted throughout the development of a system, which requires assessing the influence of design decisions on the overall risk emerging from a system on multiple levels of abstraction (e.\,g. functional, technical, software). A hazard log can serve as a baseline for each of these levels. With respect to life cycle activities, a hazard log supports an ongoing risk management after the deployment of a system. 

Taking measures to address hazards and respective hazardous events is not sufficient to show conformance with risk acceptance criteria, as this would require the ability to \emph{completely} mitigate hazardous events in all known (and unknown) scenarios. However, proving such a \emph{complete} mitigation is not only tedious but in the case of an open operational environment impossible. The concept of risk facilitates the assessment of the system with respect to acceptance criteria by providing a metric that can be applied to the entirety of hazardous events or each event individually. Therefore, risk inherited by the system of interest shall be estimated as also required by UL~4600 and ISO~Guide~51. In order to \emph{estimate risk}, several techniques are proposed in the literature \cite{noauthor_risk_2019, de_gelder_risk_2021}. Analyses based on accident databases can, for example, be carried out in simulation environments as shown by Stark~\emph{et~al.}~\cite{stark_quantifying_2019, stark_towards_2019}. Kramer~\emph{et~al.}~\cite{kramer_identification_2020} propose a risk estimation based on the probability of triggering environmental conditions potentially leading to harm with an estimated severity~\cite{kramer_identification_2020}. Such estimations can be refined over the life cycle of an Automated Driving System. 

As the basis of risk estimation, ISO~26262 defines parameters for the classification of hazardous events. Severity, exposure, and controllability allow for a subsequent risk assessment. Here, we emphasize that in the case of ISO~26262 the inference of an ASIL directly on the basis these parameters omits the step of explicit risk assessment. Nevertheless, the defined parameters enable well practiced risk management for the functional safety of conventional road vehicles (and those equipped with driver assistance systems). In contrast, for Automated Driving Systems we propose to be more rigorous in the adherence to ISO~Guide~51, as there is no standard set of sufficient safety measures specified for these systems yet. Therefore, we distinguish between \emph{determining parameters} for estimating risk and \emph{deriving (initial) risk} explicitly on the basis of severity and probability as defined in \autoref{sec:term}. This allows for a discrete classification of risk as well as calculating risk in a mathematically continuous manner. It is important to determine the initial risk without safety mechanisms as well as the residual risk actually inherited by the system of interest with the defined safety measures in place. This distinction enables the estimation of the actual risk and the determination of the necessary safety integrity.

\subsection{Risk evaluation}
The purpose of risk evaluation is to determine whether the unreasonable level of risk has been exceeded. However, determining whether tolerable risk has been exceeded comes with its own challenges: As risk acceptance criteria are given in multiple ways~(e.\,g. number of deaths per year, number of severe accidents per year, or even in general a positive balance of risk), it is not straightforward to evaluate the residual risk actually inherited by the system of interest with the accepted risk. An issue critically discussed, for example, by the \emph{German Ethics Commission on Automated and Connected Driving} is the question of weighing multiple kinds of risk. What was made clear in their report from 2017 is that \enquote{in the event of unavoidable accident situations, any distinction based on personal features (age, gender, physical or mental constitution) is strictly prohibited [-- at least from a German legislative perspective]. It is also prohibited to offset victims against one another.}~\cite{di_fabio_ethics_2017}. These ethical rules formulated by the ethics commission prohibit risk assessments that are conducted on the basis of dilemma situations (cf. the trolley problem~\cite{thomson_killing_1976}) as, for example, proposed by Geisslinger~\emph{et~al.}~\cite{geisslinger_autonomous_2021}. A question that remains open is how society handles different combinations of risk. Wachenfeld~\cite{wachenfeld_how_2017} elaborates on such issues in the context of stochastic risk evaluation. 

In order to account for approaches addressing these questions we explicitly include the step of \emph{risk modeling} as part of the risk evaluation in the Risk Management Core. Here, we define three basic activities to be conducted:

\begin{itemize}
    \item ascribing each identified risk to respective risk acceptance criteria
    \item weighing kinds of risk and considering their correspondence to one another
    \item comparing actual (residual) risk with an unreasonable level of risk
\end{itemize}

For ascribing identified risk to risk acceptance criteria, it is important to note that there can be multiple criteria. Hence, an identified risk may have to be evaluated more than once. The comparison of residual risk and accepted risk is required to determine, whether there is unreasonable risk inherited by the system of interest~(i.\,e. whether it is deemed safe). If that is the case, the safety assurance process continues with crafting a safety argumentation, which comprehensibly lays out the claims, strategies, and evidences on why the system is in fact believed to be free from unreasonable risk. On the other hand, if risk acceptance criteria are found to be violated, a certain amount of risk reduction is required, which would require the execution of further risk management iterations.

\subsection{Risk treatment} \label{sec:risk_treat}
Safety measures are determined by two factors: Firstly, the nominal demand for risk reduction shall be identified based on the difference between the residual risk and the accepted risk. Secondly, determining the initial level of risk is necessary to determine the required \emph{safety integrity}. This safety integrity also is part of the total risk reduction demand. Fowler~\cite{fowler_iec_2022} visualizes these risk budgets (cf. \autoref{fig:fowl}) by distinguishing between \enquote{minimum achievable risk}, \enquote{residual risk}, \enquote{tolerable risk}, and the risk originally emerging from the system of interest. The maximum achievable risk reduction determines the minimum achievable risk by assuming failure-free safety functions. Subsequently, some risk reduction is lost in relation to the integrity of safety functions generally being less than \qty{100}{\percent} (cf. $R(L)$ in \autoref{fig:fowl}). As safety functions add functionality to the system of interest, there can be new risk emerging from their potentially corrupt behavior (cf. $R(C)$ in \autoref{fig:fowl}). Finally, the residual risk resulting from these risk-contributing factors shall be below the unreasonable level of risk.

\Figure[ht](topskip=0pt, botskip=0pt, midskip=0pt)[width=0.9\columnwidth]{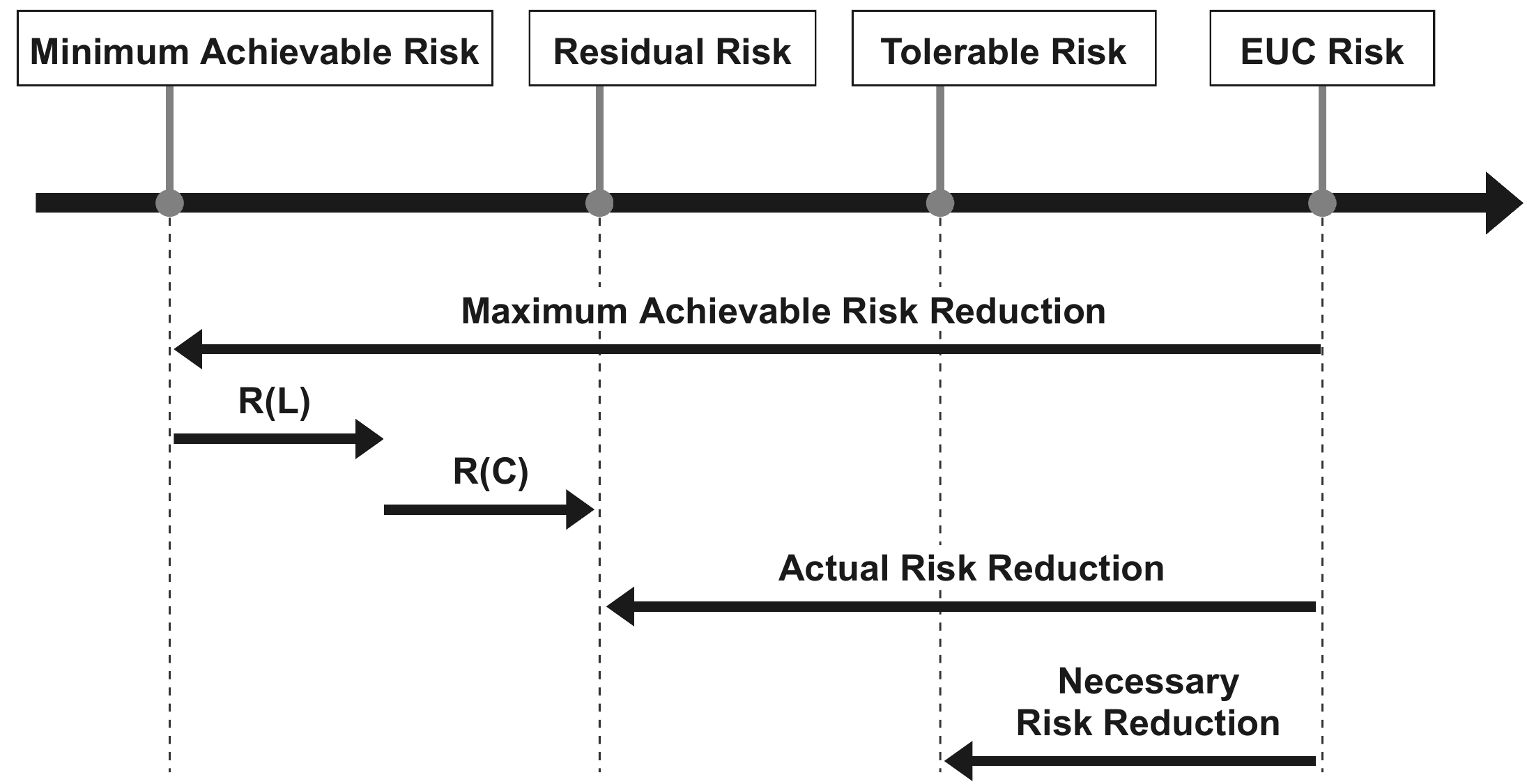}
{Fowler~\cite{fowler_iec_2022} visualizes the components of risk reduction described in IEC~61508 with respect to an equipment under control (EUC) in a one-dimensional graph. R(L) is the risk resulting from loss-type failures. R(C) refers to the risk resulting from corrupt behavior of the safety functions.\label{fig:fowl}}

In \autoref{sec:term} we defined safety integrity and risk reduction as two measures of performance~(i.\,e. a measurement) in the context of risk management. To perform risk treatment adhering to IEC~61508, safety measures~(i.\,e. preventive actions) need to be specified including both, risk reduction \emph{and} safety integrity. 
The Risk Management Core includes \emph{safety goals} as high-level safety requirements that account for the required nominal risk reduction as well as the respective safety integrity.\footnote{Note: We make this choice for the sake of compatibility to existing ISO~26262-compliant processes. Even though IEC~61508 does not introduce the concept of safety goals, the interpretation of safety goals as high-level safety requirements is not incompatible with IEC~61508.}  Subsequently, \emph{measures} shall be specified that satisfy the risk reduction and integrity needs. These measures finally influence the probability and/or severity of hazardous events and thereby contribute to the reduction of risk inherited by the system of interest. A following iteration of risk evaluation will show, whether these measures actually lead to the satisfaction of the risk acceptance criteria.

\Figure[ht](topskip=0pt, botskip=0pt, midskip=0pt)[width=2\columnwidth]{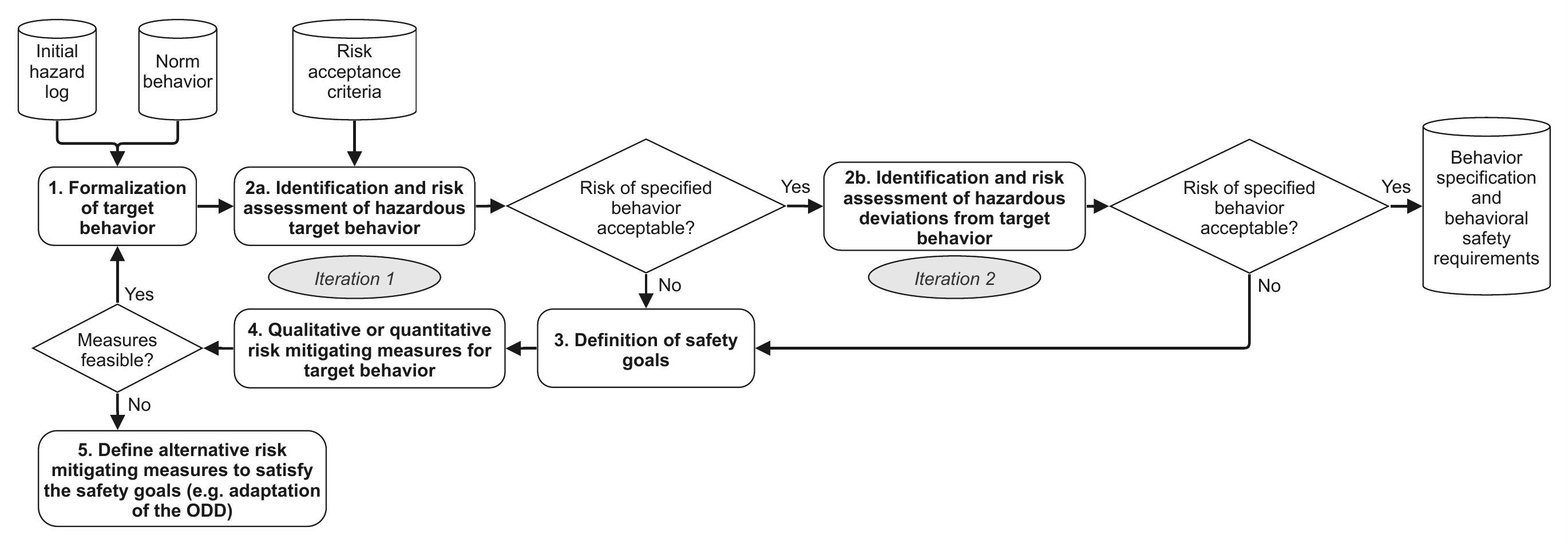}
{In order to argue for the absence of unreasonable risk in the context of the behavior specification of an Automated Driving System, a risk-based refinement of the functional specification can support the argumentation with a respective process. The result of the proposed process is a behavior specification that contains behavioral safety requirements including their required integrity.\label{fig:bsc}}

\subsection{Safety argumentation} \label{sec:safety_arg}
Once risk analysis, risk evaluation and risk treatment have been successfully executed and iterated, the determined residual risk is below the unreasonable level of risk. In the context of the approval process of automated road vehicles, this conclusion and the way it was generated, need to be transparently communicated to external stakeholders as a basis for taking an approval decision. The Risk Management Core facilitates such an explicit representation and a subsequent communication of risk.

Safety cases \cite{kelly_safety_2018} have been proposed as a means to provide an explicit representation of a structured argument, which justifies claim satisfaction based on produced pieces of evidence. According to Hawkins~\emph{et~al.}~\cite{hawkins_new_2010}, a safety case consists of two argumentation portions: Firstly, the primary argument establishes the argumentative relationship between process outputs and the claim to be supported. In the context of the Risk Management Core, this means that an explicit argumentation needs to be laid out, as to why the artifacts generated during risk analysis, evaluation and treatment support the claim ``Actual risk $<$ Unreasonable risk'' for an ADS in its operational design domain. Secondly, the confidence argument enables to document and address concerns with respect to uncertainties associated with aspects of the primary argument or supporting evidence. For instance, the estimation of (initial) risk is, in practice, not entirely objective due to being performed by safety engineers with varying experience and system knowledge \cite{khastgir_towards_2017}. Thus, to avoid the consequences of inadequately estimated risk, the confidence argument documents the concern and justifies the respective measures used to address it.

ISO~26262 \cite[Part 2, Clause 6.4.8]{noauthor_road_2018} already accepts safety cases as one way to \enquote{progressively compile the work products that are generated during the safety life cycle to support the safety argument} \cite[Part 2, Clause 6.4.8.2]{noauthor_road_2018} \enquote{for the achievement of functional safety} \cite[Part 2, Clause 6.4.8.1]{noauthor_road_2018}. To that end, since the Risk Management Core is a framework for systematically managing risk with a broader scope than functional safety and the complexity of the safety argumentation for ADS in open traffic contexts is much higher compared to systems with lower automation levels, we argue that a justifiable approval decision requires a safety case as an input explicitly documenting primary and confidence arguments for the Risk Management Core processes. Initial structuring principles for such arguments have been proposed in \cite{galbas_safeguarding_2022}.

\section{Application of the Risk Management Core to Specifying Behavior of an Automated Driving System} \label{sec:risk_bspec}
In the following, we will use the Risk Management Core to show how the proposed process framework can be applied to risk management in the context of specifying behavior of Automated Driving Systems. ISO~21448 requires an argumentation for the absence of unreasonable risk due to functional insufficiencies. Functional insufficiencies are defined as either performance or specification insufficiencies~\cite{noauthor_road_2022}. ISO~21448 describes general guidelines for identifying performance insufficiencies. However, there is no reference to a method addressing specification insufficiencies. Such insufficiencies can be caused by missing knowledge of required behavior in the operational environment \cite{stellet_formalisation_2019} and as a result can contribute to unreasonable risk inherited by the system. In this section, we propose such a method in order to contribute to a risk-based refinement of the functional specification and show the application of this method in an example scenario.

\subsection{Risk-based Refinement of the Functional Specification} \label{sec:rffs}
A behavior specification is one part of the functional specification as it is required by ISO~21448. The presented approach for a risk-based refinement of the functional specification (\autoref{fig:bsc}) requires that a method is available to derive a formal representation of target behavior.

In this context, a possible approach that allows making design assumptions and decisions in a behavior specification explicit was proposed by Beck~\emph{et~al.}~\cite{beck_phenomenon-signal_2022, beck_phanomen-signal-modell_2021}. The \enquote{Phenomenon-Signal Model} (PSM) provides means for explicitly representing facts and rules that constitute the target behavior of an Automated Driving System in road traffic. Additionally, Salem~\emph{et~al.}~\cite{salem_beitrag_2022} and Beck~\emph{et~al.}~\cite{beck_contributions_2022} present the \enquote{Semantic Norm Behavior Analysis} as an approach to traceably derive such facts and rules based on stakeholder concerns, more specifically, legal concerns derived from the German traffic code. The combination of facts and rules in a graph-representation yields a formal specification of ADS target behavior. 

For our purposes, we follow Beck~\emph{et~al.}~\cite{beck_phenomenon-signal_2022} and base the formal target behavior description (cf. \autoref{fig:bsc}, Step 1) on a collection of stakeholder concerns (norm behavior, according to \cite{beck_phenomenon-signal_2022}) and an initial representation of possible hazards (hazard log).
The first step of \emph{formalizing target behavior} shall support a risk-based refinement of the functional specification by enabling a representation of hazardous behavior~(i.\,e. behavior leading to hazardous events). Furthermore, triggering conditions and functional insufficiencies need to be represented in order to address them explicitly with risk reduction measures (qualitatively as well as quantitatively). The formal, graph-based representation of target behavior provided by the Phenomenon-Signal Model (PSM) satisfies each of these requirements, as the representation relies on facts and rules which can be derived from required risk reduction measures. \enquote{In a similar way as the legal sciences, which essentially determine behavioural [sic!] norms, link legal consequences and offenses in laws, facts are all those established circumstances which justify the application of a rule, that is, represent the ‘IF’ condition.}~\cite{beck_phenomenon-signal_2022}

Step 2a (cf. \autoref{fig:bsc}) addresses the \emph{identification of hazardous target behavior}. If we consider target behavior to be a set $\mathcal{T}$ and hazardous behavior to be a set $\mathcal{H}$, then this step aims to identify and analyze the behavior-related risk $r_{\mathrm{beh},\mathrm{t}}$ emerging from $\mathcal{T} \cap \mathcal{H}$. Subsequently, the behavior-related risk is a mapping $f$ given as $r_{\mathrm{beh},\mathrm{t}} = f(\mathcal{T} \cap \mathcal{H})$ which is evaluated by comparing the (estimated) actual risk to the acceptable level of risk $r_\mathrm{rac}$. Note, that such a scalar evaluation is meant to illustrate a comparison, while we acknowledge the potentially higher dimensionality of actual and accepted risks.

In the case of $r_{\mathrm{beh},\mathrm{t}} > r_\mathrm{rac}$, risk reduction is required. \emph{Safety goals are defined} (cf. \autoref{fig:bsc}, Step 3) to capture the risk reduction needs qualitatively and quantitatively. Quantities shall be given for both nominal risk reduction and safety integrity. Note that this step does not necessarily require a definition of new safety goals but can also be performed by refining existing safety goals.

Safety goals shall be satisfied by \emph{specifying safety measures} (cf. \autoref{fig:bsc}, Step 4). For defining safety measures on the level of the behavior specification, we introduce the concept of behavioral safety requirements. Behavioral safety requirements are a specialization of functional requirements that specify behavior of a system of systems in a scenario-specific context. In contrast, traditional functional safety requirements (especially in the context of ISO~26262) refer to an item in its entire operational environment. We allocate behavioral safety requirements to architectural elements using the concept of capabilities. The use of capabilities as architectural elements being required for a certain target behavior is elaborated in previous work \cite{reschka_ability_2015,nolte_towards_2017,bagschik_systems_2018}. Stolte~\emph{et~al.}~\cite{stolte_towards_2020} adhered to the terminology of ISO~26262 by only specifying functional safety requirements and allocating these to capabilities. However, we argue that in order to conform with requirements from both ISO~26262 and ISO~21448, a distinction between functional safety requirements and behavioral safety requirements can be useful. For functional safety, an argument shall be provided, laying out why the absence of unreasonable risk due to hazards caused by malfunctioning behavior of electrical and/or electronic (E/E) systems is achieved. However, for SOTIF the absence of unreasonable risk due to hazards resulting from functional insufficiencies of the intended functionality or its implementation shall be argued. Since a specification of safe target behavior is assumed in the context of functional safety, SOTIF-related activities shall (among others) lead to a specification of safe target behavior. Behavioral safety requirements could thus support the vehicle-level SOTIF strategy (VLSS), which solely includes vehicle-level requirements for the intended functionality, especially by explicitly adding risk reduction and safety integrity as associated information. Additionally, a rigorous description of target behavior could support activities in the functional safety life cycle such as a systematic hazard identification \cite{kramer_identification_2020, graubohm_towards_2020} and a subsequent specification of safety goals \cite{stolte_towards_2020}.

The scope of our example application of the Risk Management Core only addresses the behavior specification. Hence, (e.\,g. economically or technologically) infeasible risk reduction measures could be defined in the first iteration. If that is the case, \emph{alternative measures} to adapting target behavior can be specified (cf. \autoref{fig:bsc}, Step 5). Such measures could, for example, include modifications of the operational design domain.

Finally, \emph{iteration~1} is completed by integrating the specified risk reduction measures into the behavior specification, which is again followed by a risk analysis of the target behavior. Iteration~1 thereby deals with reducing $r_{\mathrm{beh},\mathrm{t}} = f(\mathcal{T} \cap \mathcal{H})$ to an acceptable amount.

Assuming that there is a point at which the residual risk emerging from the target behavior of the system of interest reaches an acceptable amount, \emph{iteration~2} begins. As it is defined in ISO~21448, triggering conditions that can lead to deviations from the target behavior shall be analyzed (cf. \autoref{fig:bsc}, Step 2b). If the deviations are found to be hazardous, the risk emerging from such kind of behavior $r_{\mathrm{beh},\mathrm{d}} = f(\mathcal{H} \setminus \mathcal{T})$ needs to be quantified. In order to successfully exit iteration~2 risk emerging from the target behavior has to be below the acceptable level of risk $r_{\mathrm{beh},\mathrm{t}} + r_{\mathrm{beh},\mathrm{d}} < r_\mathrm{rac}$. If the risk assessment shows that at least one risk acceptance criterion is violated, further risk reduction is necessary. Risk emerging from specified safety mechanisms (cf. Fowler~\cite{fowler_iec_2022}) will necessarily be analyzed by reentering iteration~1. In case all risk acceptance criteria are satisfied by the behavior specification -- including risk reduction and safety integrity -- a refined behavior specification can be provided.

\subsection{Example of a Refined Behavior Specification}
In this section, we show how the risk-based refinement of a functional specification -- described in \autoref{sec:rffs} -- can be applied in an example scenario. In this scenario\footnote{This example scenario serves as a means to show traceability of the approaches developed within the \emph{PEGASUS Family} research project \emph{VVMethods}.} we consider an urban T-crossing (\autoref{fig:scen}). The automated ego vehicle approaches the crossing from the left-hand side of the figure. A crosswalk is located to the left of the T-crossing. The crosswalk is signposted by markings and traffic signs. A pedestrian intends to use the crosswalk. At this abstraction level, the scenario is not further parameterized, which makes it a functional scenario as defined by Menzel~\emph{et~al.}~\cite{menzel_scenarios_2018}.

\Figure[ht](topskip=0pt, botskip=0pt, midskip=0pt)[width=0.8\columnwidth]{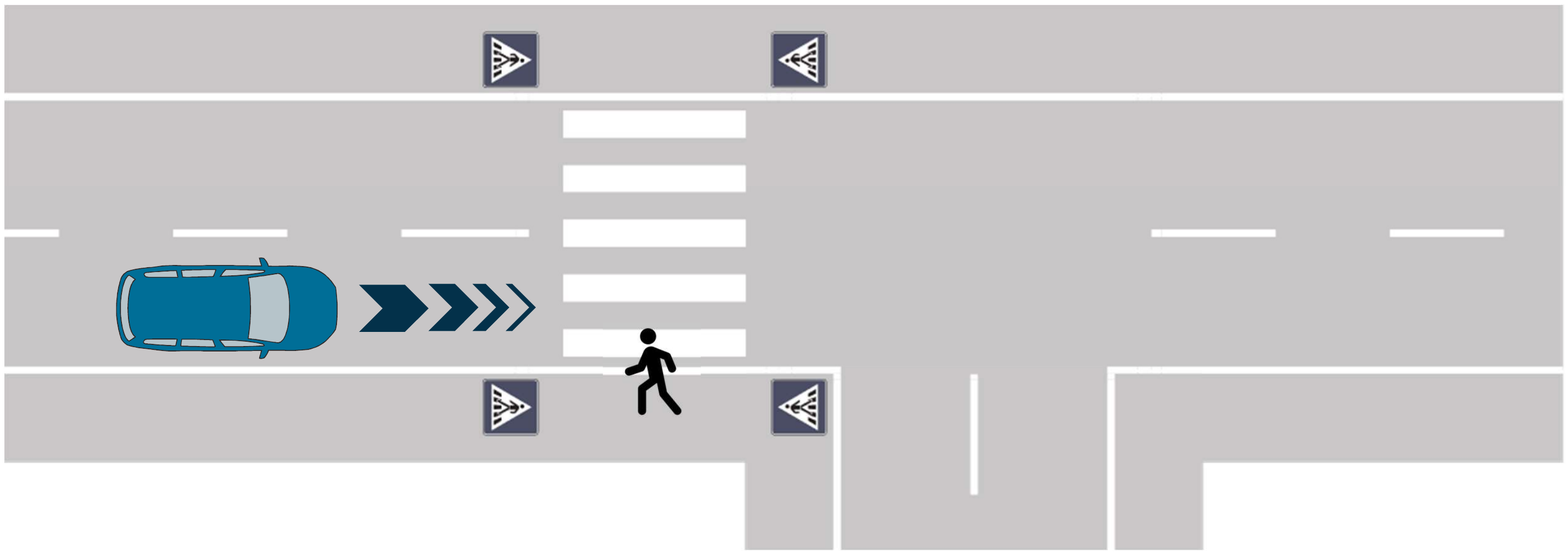}
{In this example, an urban T-crossing is considered. A pedestrian intends to use a crosswalk, which is signposted by respective markings and traffic signs. The automated ego vehicle enters the scenario from the left side of the figure.\label{fig:scen}}

Based on the Phenomenon-Signal Model (PSM)~\cite{beck_phenomenon-signal_2022, beck_phanomen-signal-modell_2021} a formal specification of target behavior is constituted by facts and rules. A semi-formal representation of the knowledge that is captured in a Phenomenon-Signal Model (PSM) is depicted in \autoref{fig:vspek}. We assume the initial target behavior to be given in the scope of this paper after having performed an example analysis.

There are multiple assumptions related to such an analysis. Here, we do not focus on formulating them explicitly. We rather use the following example facts and rules to show the process described in \autoref{sec:rffs}. As an input we define the necessary knowledge about norm behavior. Such knowledge is in our case given by performing a Semantic Norm Behavior Analysis~\cite{salem_beitrag_2022, beck_contributions_2022} of the example scenario and parts of the German traffic code (sec.~26 par.~1). Numbers for the markings and signs are given as defined in the German traffic code.\\

\noindent
\textbf{Facts:}
\begin{itemize}
    \item crosswalk marking (sign number~293) is detected
    \item crosswalk sign (sign number~350) is detected
    \item crosswalk is detected
    \item pedestrian is detected
    \item pedestrian position near crosswalk
    \item crossing intention of pedestrian is detected
    \item ego position near crosswalk
    \item stop at the crosswalk
\end{itemize}
\vspace{3mm}
\noindent
\textbf{Rules:}
\begin{itemize}
    \item \textbf{If} crosswalk marking (sign number~293) \textit{and} crosswalk sign (sign number~350) \textbf{then} valid crosswalk
    \item \textbf{If} pedestrian position near crosswalk \textbf{then} pedestrian intends to use it
    \item \textbf{If} pedestrian intends to use a crosswalk \textit{and} the ego approaches that crosswalk \textbf{then} the ego shall enable the pedestrian to do so by the ego vehicle stopping at the crosswalk
\end{itemize}

\Figure[ht](topskip=0pt, botskip=0pt, midskip=0pt)[width=2\columnwidth]{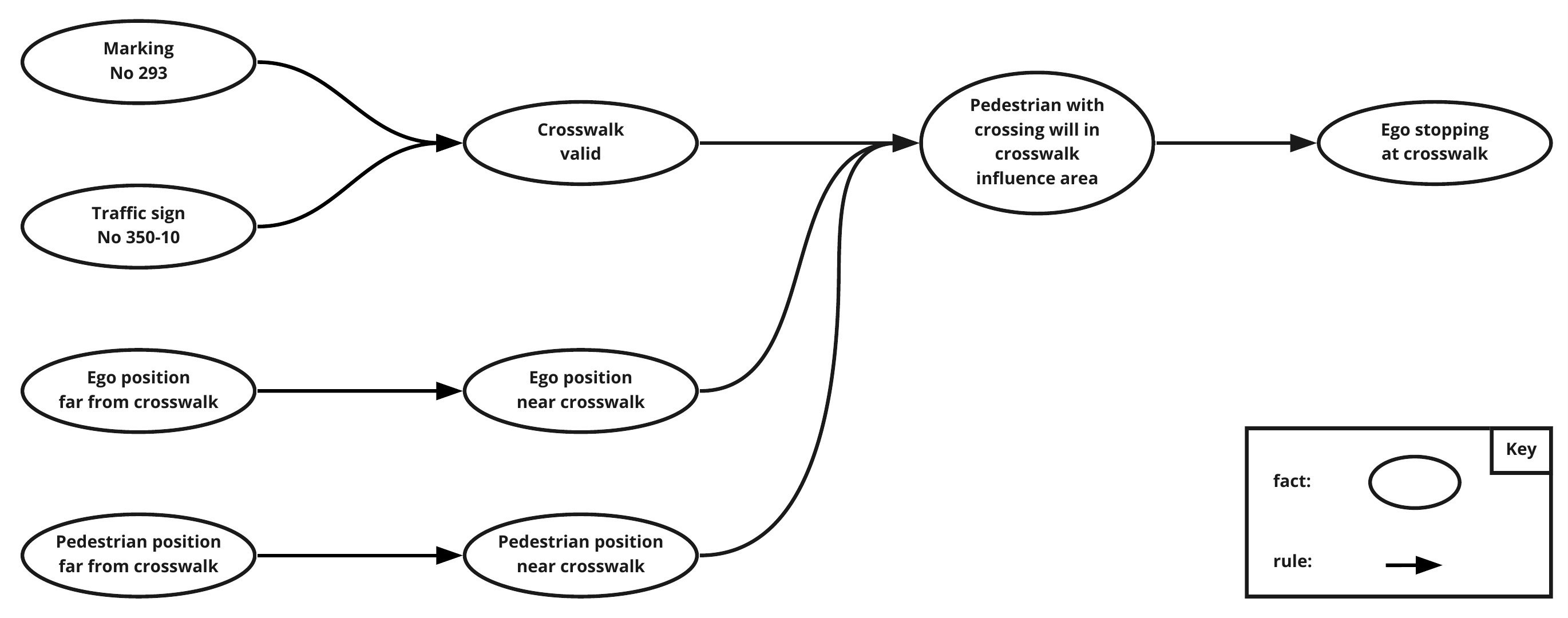}
{The formal specification of target behavior (based on facts an rules that are applied to a scenario) requires an explainable representation. Hence, we utilize causal graphs to capture the respective knowledge in human-readable format.\label{fig:vspek}}

\vspace{3mm}
Furthermore, an initial hazard log is given. Hazard sources are an input to the Risk Management Core. In an example hazard analysis, we assume that one hazard in the use case of crossing a T-crossing with only the two mentioned actors has been previously identified.\\

\noindent
\textbf{Hazard:}
\begin{itemize}
    \item Road vehicle collides with a vulnerable road user (VRU) at a pedestrian crossing, resulting in potential injury of the vulnerable road user\footnote{We follow the ontology in \autoref{fig:risk_assess} and formulate a rather specific hazard here -- including the agents (road vehicle, vulnerable road user), the hazardous event (collision) the possible harm (injury), as well as use case-specific context (at a pedestrian crossing) in the hazard formulation.}.
\end{itemize}

\vspace{3mm}
Following the process specified in \autoref{sec:rffs}, we search for hazardous target behavior within iteration~1 (\autoref{fig:bsc}). In the example scenario, the facts and rules from the target behavior specification state that the ego vehicle is supposed to stop at the crosswalk. 

For the identification of hazardous events that lead to harm (in terms of physical injury), we only consider the external behavior of the vehicle in traffic to be relevant. Thus, we analyze, whether a specification that demands stopping at a crosswalk can potentially cause physical injuries to the pedestrian. In our example, this is not the case as the behavior specification for this example scenario does not include hazardous behavior. A slight variation of the scenario, however, gives an example of how the given behavior specification can lead to harm: If we assume that the pedestrian does not walk directly on the crosswalk but instead crosses the road, for example, between the ego and the crosswalk, the target behavior remains unchanged (i.\,e. the ego vehicle continues following its target speed). This is, because in the modified scenario where the pedestrian crosses the road far from the crosswalk, the fact \emph{pedestrian position near crosswalk} is not satisfied. Hence no \emph{stopping maneuver} is required~-- at least according to the formulated behavior specification. Given that the pedestrian crosses the road as specified in the scenario variation, there is a potential contact between the two agents. This leads to the following hazardous event.

\vspace{6mm}

\noindent
\textbf{Hazardous event:}
\begin{itemize}
    \item Road vehicle collides with a pedestrian in front of a crosswalk.
\end{itemize}

\vspace{3mm}
This hazardous event is an instance of the specified hazard in the example scenario. Whereas the vulnerable road user does not need to be further specified in a use case description, further details have to be provided in scenario-specific context. Hence, the hazardous event refers to the pedestrian as an instance of a vulnerable road user. Such a hazardous event would be the result of a specification insufficiency, where pedestrians with irregular crossing behavior were not considered during the formalization of target behavior.
 
Assessing risk without meaningful data is challenging~-- particularly with regard to information about the severity of harm potentially resulting from a hazardous event. Since the focus of this work is arguing for a risk management methodology, we do not attempt to provide a valid risk quantification. However, we give an example of a potential outcome of a risk assessment.

In order to estimate the risk resulting from the identified hazardous event, we make the following assumptions: A fleet of 1,000 automated people-mover vans is operated in the Berlin city area for 365 days per year. As the vehicles in the fleet are defined to be electric vehicles, we assume 22 hours of operation per day to account for charging and maintenance. Per van and year we assume three near misses of pedestrians at crosswalks where every 3,000th is a fatal collision. Therefore, the time interval between two events with pedestrian fatalities at crosswalks in our example is one year. Within the scenario we assume that the occurrence of the hazardous event inevitably leads to harm (e.g. no controllability by the involved agents).

An estimate of the severity of harm resulting from a collision between a road vehicle and a pedestrian would require a parametrization of their behavior. In the scope of a functional scenario we make further assumptions. In Germany, crosswalks are only permitted within built-up areas. In such environments the usual speed limit is \qty{50}{\kilo\metre\per\hour}. A collision under these general circumstances could lead to \emph{life-threatening injuries} whereas survival is uncertain (assuming run over, equivalent to an S3 classification in ISO~26262 \cite{noauthor_considerations_2015}). Please note that we only quantify risk based on rough estimations. Logical and concrete scenarios~\cite{menzel_scenarios_2018} would, for example, be required for a more detailed assessment, to provide information about the masses and speeds of the involved agents.

As a result of our assumptions we represent an estimation of the actual risk by first calculating the operating hours per year. Subsequently, the number of fatalities at crosswalks per operating hour can be calculated based on our estimation of one crosswalk fatality per year for the entire fleet.

\begin{equation}
    \SI{1000}{Vans} \cdot \frac{\SI{22}{h}}{\si{\day}} \cdot \frac{\SI{365}{\day}}{\si{\year}}
    \approx \SI[per-mode = fraction,exponent-product = \cdot]{8d6}{\hour\per\year} 
\end{equation}

\begin{equation}
    \frac{\SI[per-mode = fraction]{1}{\fatality\per\year}}{\SI[per-mode = fraction,exponent-product = \cdot]{8d6}{\hour\per\year}} \approx \SI[per-mode = fraction,exponent-product = \cdot]{1.25d-7}{\fatalities\per\hour}
\end{equation}

For a risk-based refinement of the functional specification, which is based on the Risk Management Core, we defined that the evaluation of target behavior against risk acceptance criteria is required. As for the step of risk assessment, we set the focus on applying our methodology rather than identifying valid risk acceptance criteria. One risk acceptance criterion defined by the \emph{German Ethics Commission on Automated and Connected Driving} is the \emph{positive risk balance}~\cite{di_fabio_ethics_2017}. That means the introduction of an Automated Driving System should lead to a risk below the risk resulting from human driving. We assume this risk acceptance criterion to be the only relevant one for our example. 

In order to represent the accepted risk that satisfies the risk acceptance criterion in this example, we draw from accident data from current human driving. We assume that this data covers the same operational design domain (ODD) that we used for our estimation of the actual risk in order to compare the two values.

According to the Berlin senate administration~\cite{noauthor_strasenverkehrszahlung_2021}, the annual mileage on Berlin city roads is \SI[exponent-product = \cdot]{8.623d6}{\kilo\metre\per\year}, average speed is $\approx \SI{24}{\kilo\metre\per\hour}$~\cite{noauthor_europaische_2008}.
In the period from 2006 to 2011, only one accident with fatally injured pedestrian was recorded at crosswalks in Berlin~\cite{ortlepp_untersuchungen_2013}. As a result, we estimate the total driving hours in Berlin per year and subsequently quantify the frequency of a fatality at a crosswalk per driving hour in current Berlin traffic based on our assumptions.

\begin{equation}
    \frac{\SI[per-mode = fraction,exponent-product = \cdot]{8.623d6}{\kilo\metre\per\year}}{\SI[per-mode = fraction]{24}{\kilo\metre\per\hour}} \approx \SI[per-mode = fraction,exponent-product = \cdot]{3.59d8}{\hour\per\year}
\end{equation}

\begin{equation}
    \frac{\SI[per-mode = fraction,parse-numbers = false]{\frac{1}{6}}{\fatalities\per\year}}{\SI[per-mode = fraction,exponent-product = \cdot]{3.59d8}{\hour\per\year}} \approx \SI[per-mode = fraction,exponent-product = \cdot]{4.64d-10}{\fatalities\per\hour}
\end{equation}

Since the comparison of our representations of the estimated actual risk and the accepted risk of the hazardous event suggests that the threshold for the accepted risk is exceeded, the risk acceptance criterion would be violated. Hence, we need to define safety goals and subsequently specify safety requirements, which we can attribute with the necessary risk reduction. Comparing this approach with the activities described in ISO~26262, the safety goals could already be specified after hazard identification has been conducted. Additionally, legacy knowledge could be used to specify safety goals earlier in the safety life cycle. The Risk Management Core explicitly accounts for this order of proceeding. In this paper, safety goals are specified after the necessary risk reduction has been identified in order to emphasize the contribution of this work. Note, that the example safety goal refers to both, the considered hazard and the hazardous event in order to find a suitable level of abstraction (according to the hazard) and account for necessary risk reduction based on the assessment of the hazardous event. The risk assessment in our example includes assumptions about the safety integrity that is part of the necessary risk reduction required for the safety goals and behavioral safety requirements. In \autoref{sec:rmc} we elaborated that safety integrity depends on the initial risk determined for the system of interest. IEC~61508 provides guidelines on how to define safety integrity (levels) based on such an initial risk.

\vspace{3mm}

\noindent
\textbf{Safety Goal:}
\begin{itemize}
    \item Prevent collision between a road vehicle and a vulnerable road user at crosswalks.
\end{itemize}

\vspace{3mm}
With respect to our example safety goal a safety measure would be to include pedestrians showing irregular crossing behavior at crosswalks in the formalization of target behavior and define facts and rules, which capture their crossing intention. We further assume that such a safety measure\footnote{Note that the safety measures are based on rough risk estimations that can be made in a functional scenario.
However, even based on these rough estimates, it becomes obvious that first safety measures in terms of modifications to the behavior specification can be defined. More detailed risk assessments are necessary, for example, in order to specify and implement technical safety measures. Further details can be provided by considering logical and concrete scenarios.} is deemed feasible as indicated by state of the art pedestrian prediction algorithms~\cite{ridel_literature_2018}.

\vspace{12mm}

\noindent
\textbf{Behavioral Safety Requirements:}
\begin{itemize}
    \item If a crosswalk is detected, detect pedestrians crossing intention at crosswalks reliably.
\end{itemize}

\vspace{3mm}
To conclude iteration~1 (\autoref{fig:bsc}), we would have to conduct a second semantic norm behavior analysis in order to extract relevant facts and rules dealing with reasonably foreseeable irregular behavior of pedestrians at crosswalks, e.\,g. based on court cases. Additionally, helpful guidelines on assumptions with respect to other road users can be found in IEEE 2846~\cite{noauthor_ieee_2022}. For our example scenario, we specify that the crossing intention of pedestrians should not only be based on their position near a crosswalk, but that their crossing intention should be inferred based on the current traffic context. That means the ego vehicle shall detect them, infer their intention to cross and stop at the crosswalk to enable them crossing it.

In iteration~1 (\autoref{fig:bsc}) no deviations from the specified behavior are considered. As a result of including the specified safety measures in the behavior specification, the identified hazardous event would effectively not occur in our scenario catalog (of two scenarios). Note that we neglect uncertainties in the preceding process of specifying all relevant scenarios and the risk emerging from insufficiencies regarding their completeness.

Entering iteration~2 (\autoref{fig:bsc}), we need to assess the risk resulting from deviations from target behavior. Such deviations can, for example, be identified as described by Kramer~\emph{et~al.}~\cite{kramer_identification_2020}. We presented a modified application of their proposed method by applying guide words such as \emph{not}, \emph{early}, or \emph{late} in~\cite{salem_maneuver-based_2022}. One hazardous event that can be identified, which results from a deviation of the specified target behavior is the ego vehicle not stopping at the crosswalk. This could be caused by a missing detection of the pedestrian attempting to use the crosswalk. For the sake of this example, we do not conduct an additional estimation of risk, risk evaluation and treatment as it was already illustrated for iteration~1 (\autoref{fig:bsc}).

Finally, the completion of iteration~2 (\autoref{fig:bsc}) leads to a refined behavior specification. Such a specification should be consistent with following applications of the Risk Management Core~(e.\,g. risk-based refinement of the functional system architecture) in order to avoid conflicting assumptions during risk management.
    
\section{Evaluation} \label{sec:eval}
In \autoref{sec:rmc}, we proposed the Risk Management Core as a process framework that supports iterative risk management~(i.\,e. risk analysis, risk evaluation, and risk treatment) for Automated Driving Systems. Subsequently, we applied the Risk Management Core to the task of specifying safe behavior in \autoref{sec:risk_bspec} and proposed a risk-based refinement of the functional specification. To conclude, we will verify whether the requirements we elicited for risk management in the context of automated driving in \autoref{sec:req} are met by the Risk Management Core process framework.

Requirement~\ref{req1} demands the identification of hazardous events. The Risk Management Core satisfies this requirement by specifying that scenarios are analyzed with regard to the instantiation of hazards in a specific context. An identification of all hazardous events depends on the completeness of the available scenario catalog. Hence, we do not claim that the Risk Management Core enables such kind of completeness. However, by providing a systematic approach to manage risk emerging from known hazardous events it contributes to the explicit representation of known risk.

Furthermore, the assessment of the risk emerging from the identified hazardous events (Requirement~\ref{req2}) shall be enabled. In the Risk Management Core, the step of determining parameters supports both, a continuous and a discrete risk assessment. These parameters are traceably assigned to the severity of harm (Requirements~\ref{req3} and~\ref{req4}) and the probability of the occurrence of that harm (Requirement~\ref{req5}). 

Deriving the initial and~-- in later iterations the deviating~-- actual risk of the hazardous events is the subsequent step in the Risk Management Core (Requirement~\ref{req2}). In this context, so far, we have only conducted analyses based on functional scenarios, which leads to rough estimates of initial and actual risk. From a methodological point of view, however, the extension to logical and concrete scenarios is possible. 

In order for risk management activities to have a defined termination criterion, the accepted level of risk is determined based on risk acceptance criteria (Requirement~\ref{req11}). The public availability of such criteria is only partially given. ISO~21448 mentions criteria such as MEM (minimum endogenous mortality) or ALARP (as low as reasonably practicable). However, it is still an open question how these criteria can be applied to automated driving and how they relate to criteria specified by the German Ethics Commission (e.\,g. the positive risk balance).

As a result of the actual risk and the accepted risk being available, risk evaluation is performed, in order to determine the necessary risk reduction and the required safety integrity (Requirement~\ref{req15}). This step involves the step of risk modeling, which we only outlined. Challenges in this task include weighing multiple kinds of risk in a way that is accepted by society. While the Risk Management Core does not contribute to this task specifically, it provides a framework to assess and treat risk as consequence of the involved trade-offs.

The risk reduction demand is captured by specifying safety goals (Requirement~\ref{req12}) that should lead to an acceptable mitigation of risk emerging from the identified hazardous events. In order to achieve the safety goals, measures are specified (Requirements~\ref{req8}, \ref{req9}, and~\ref{req10}) that mitigate the risk to the desired amount. Finally, the safety goals are not only linked to hazardous events by specifying measures but they are also traceable to the known hazards (Requirement~\ref{req13}) that are addressed by these safety goals. Such traces are documented in a hazard log (Requirement~\ref{req14}). In contrast to existing systems, for Automated Driving Systems there are no best practices on measures available that lead to a reasonable level of risk. As soon as such best practices are established in the automotive industry~-- comparable to measures specified in ISO~26262~-- the current effort for an explicit derivation of risk reduction based on acceptance criteria might become more straightforward.

Finally, we assess the Risk Management Core with respect to its limitations.

First and foremost, the Risk Management Core is a framework that shall enable risk management. For its practical application, valid risk acceptance criteria are required. Achieving consensus for these criteria remains a societal challenge including both, interdisciplinary communication with inconsistent terminology and negotiations about valid criteria.

Second, we highlighted necessary inputs for the Risk Management Core besides risk acceptance criteria. Hazard sources and scenarios (ideally including parameters with probability distributions) are indispensable to conduct risk analyses. Each of these inputs for risk analysis and risk evaluation needs to be updated throughout the life cycle of an Automated Driving System in order for the results of the Risk Management Core to remain valid after deployment.

Furthermore, risk assessment (at least partially) requires quantified risk estimation and assessment. 
In order to assess compliance of an Automated Driving System with the defined risk acceptance criteria, actual and accepted risks need to be compared. However, the Risk Management Core itself does not answer the questions that arise when a straightforward comparison of actual and accepted risks is not feasible. In these cases there is a need for approaches to conduct risk modeling in the context of automated driving. Note, that the term \enquote{straightforward} does only refer to the comparison of actual and accepted risks if they are of the same dimensionality. We acknowledge that quantitatively estimating risk remains a challenge involving, for example, representative data on scenario probability, detailed knowledge about failure rates and performance limitations.

Finally, we emphasize that the Risk Management Core focuses on representing risk and does not explicitly account for security vulnerabilities to be potential sources of harm. While the challenges of ensuring security is key in the context of cyber-physical systems (also when considering the collection of fleet data), potential effects of security vulnerabilities to the safety of an Automated Driving System (i.e. the consideration of security vulnerabilities as risk sources) is not discussed as part of this work. For a more in-depth discussion on security effects on safety we refer the reader to \cite{girdhar_post-accident_2022}.
    
\section{Conclusion and Outlook}
In this paper, we discussed the need for an explicit representation of risk in the context of automated driving. While ISO~21448 requires the definition of acceptance criteria, no guidelines are provided on how the absence of unreasonable risk can explicitly be verified. ISO~26262 solely provides guidelines for implicit risk management that~-- without further measures~-- could potentially lead to a lack of acceptance of Automated Driving Systems in society. Explicit assessment and treatment of risk, however, has the potential to explicitly include various stakeholder perspectives on safety and accepted risk. Hence, it could provide a basis for targeted risk communication and hence provide a basis for improving societal acceptance of automated-driving-related technology. Furthermore, managing risk in addition to the treatment of worst-case hazardous events across different scenarios supports a more dedicated implementation of safety measures that address unreasonable risk in a specific context.

As means to support the endeavor of explicit risk management, we elicited requirements for risk management in the context of automated driving (\autoref{sec:req}), proposed an ontology that is based on the terminology used in multiple (automotive) safety standards (\autoref{sec:term}), elaborated the Risk Management Core as a risk management framework (\autoref{sec:rmc}), applied this process framework to the task of behavior specification of Automated Driving Systems in an example scenario (\autoref{sec:risk_bspec}) and analyzed whether the requirements are met by the proposed process framework (\autoref{sec:eval}).

Future research could contribute to explicit risk management by applying the Risk Management Core to further tasks of assuring the safety of Automated Driving Systems, for example, by detailing the approach in the context of functional safety. Furthermore, the relation between the presented work and existing standards can be further detailed. 
While in this article we have elicited requirements based on safety standards that are widely adopted, an explicit integration into existing safety standards still needs to be conducted. Additionally, integrating the Risk Management Core into processes of involved stakeholders, for example, car manufacturers or type approval authorities is an open challenge that needs to be addressed by the respective stakeholder\footnote{
Note, that this work was performed within a the publicly funded project VVMethods including multiple stakeholder from the German automotive industry. This context requires the approaches to remain neutral with respect to factors relevant for economic competitiveness -- including integration into existing processes of the project partners.
}. In case the Risk Management Core is adopted in such processes,
it remains an open question which risk acceptance criteria are valid for Automated Driving Systems and will require negotiations on a societal level. Additionally, we will focus on showing a risk-based refinement of the functional specification by not only conceptually but formally integrating the Risk Management Core and the Phenomenon-Signal Model (PSM) in the future.

\section*{Acknowledgment}
We thank our colleagues Hans Nikolaus Beck, Hagen Böhmert, Matthias Rauschenbach, Stefan Heiss, Marvin Loba, and Markus Steimle for contributing to the discussions regarding this work. Additionally, we thank our partners in the \enquote{PEGASUS Family} project \enquote{Verifikations- und Validierungsmethoden automatisierter Fahrzeuge im urbanen Umfeld} for their feedback.

\bibliographystyle{IEEEtran}
\bibliography{Riscore}

\begin{IEEEbiography}[{\includegraphics[width=1in,clip,keepaspectratio]{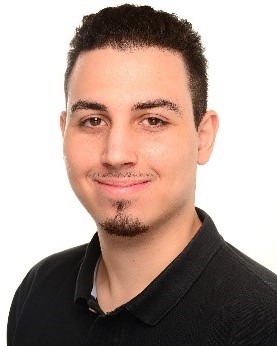}}]{Nayel Fabian Salem} received the B.Sc. degree in mechanical engineering from Technische Universit\"at Berlin (2018), Berlin, Germany, and the M.Sc. degree in electronic systems engineering from
the Technische Universität Braunschweig (2020), Braunschweig, Germany.

Since 2021, he has been a Research Associate pursuing his PhD at the Institute of Control Engineering, Technische Universit\"at Braunschweig. His main research interests include safety assurance of automated vehicles, focusing on safety issues regarding behavior specification.
\end{IEEEbiography}

\begin{IEEEbiography}[{\includegraphics[width=1in,clip,keepaspectratio]{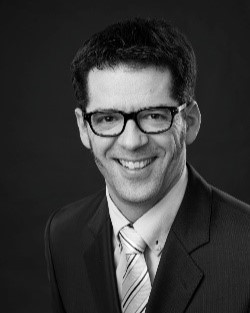}}]{Thomas Kirschbaum} received the Dipl.-Ing. (BA) degree in communications engineering from Berufsakademie Stuttgart in 1998. He has been employed at Robert Bosch GmbH since 1995. As Senior Expert System Safety he is currently contributing his many years of experience in the field of safety to the standardization of highly automated road vehicles (ISO/TR 4804, ISO/TS 5083) and to the publicly funded project VVMethoden. Previously, Thomas was responsible for the safety aspects of railway projects, including his role as a recognized expert for the German Federal Railway Authority. Since the emergence of the ISO 26262 safety standard and its in-house roll-out in 2009-2013, Thomas has had responsibility for various safety topics. In addition, his professional career includes projects in software and hardware development as well as responsibility in quality assurance and product management.
\end{IEEEbiography}

\begin{IEEEbiography}[{\includegraphics[width=1in,clip,keepaspectratio]{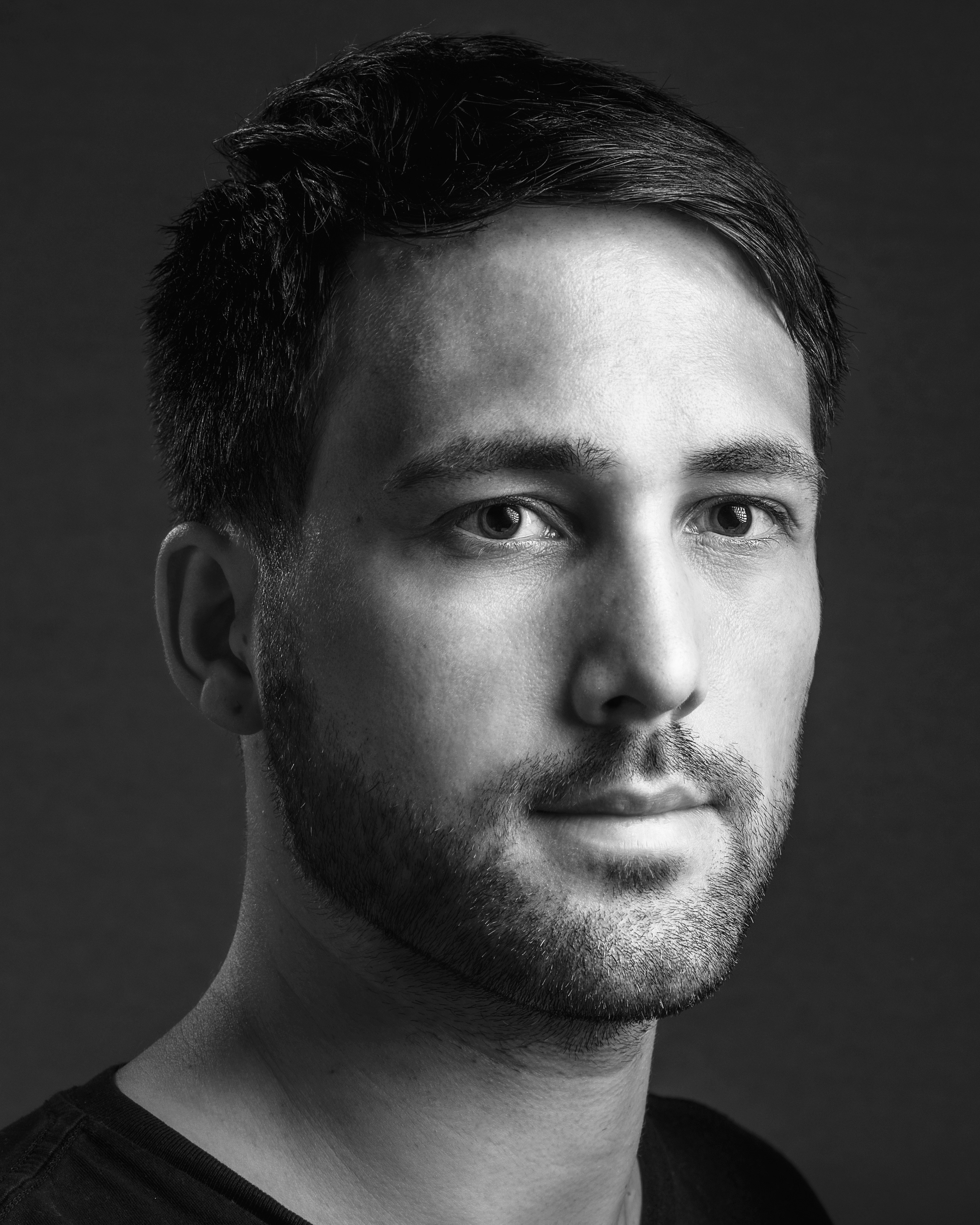}}]{Marcus Nolte} works as a Research Associate at the Institute of Control Engineering at TU Braunschweig since 2014 and is currently pursuing his PhD. He received his Bachelor and Master of Science in electrical engineering from TU Braunschweig in 2011 and 2014. His main research interest is self- and risk-aware and motion planning for automated vehicles.
\end{IEEEbiography}

\begin{IEEEbiography}[{\includegraphics[width=1in,clip,keepaspectratio]{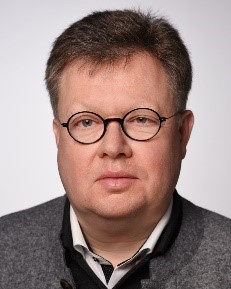}}]{Christian Lalitsch-Schneider} received a Dipl.-Ing. (Univ.) degree in manufacturing engineering at the Friedrich-Alexander-Universität Erlangen-Nürnberg in 1998. Christian joined ZF Friedrichshafen AG as a systems architect in the central research and development department at the beginning of 2019. The focus of his tasks is the daily use of systems engineering in the concept development of mechatronic and cyber-physical systems as well as in the skills development of colleagues in the field of model-based systems engineering. Christian has been a certified systems engineer by conviction for many years, because he sees systems engineering as the guiding principle for exceeding customer needs. In his role as systems architect he is currently contributing to the publicly funded project VVMethoden. In the German Chapter of INCOSE Christian is the honorary project leader for the organization of the online seminar series.
\end{IEEEbiography}

\begin{IEEEbiography}[{\includegraphics[width=1in,clip,keepaspectratio]{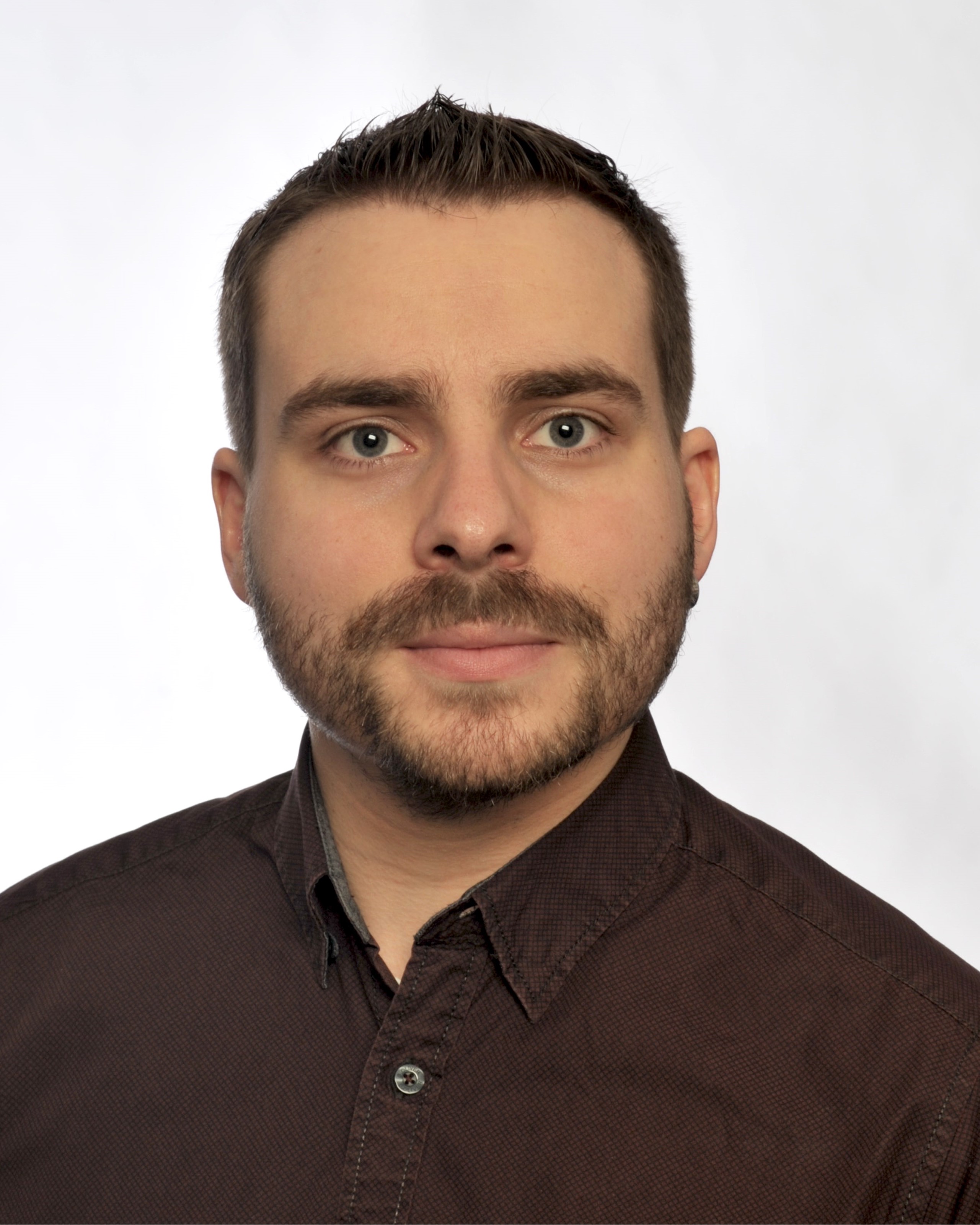}}]{Robert Graubohm} received the B.Sc. (2013) and M.Sc. (2016) degree in
industrial engineering in the field mechanical engineering from Technische Universität Braunschweig,
Braunschweig, Germany, and the M.B.A. (2015) degree
from the University of Rhode Island, Kingston, RI, USA. He is currently a Research
Associate with the Institute of Control Engineering,
Technische Universität Braunschweig. His main research interests include development processes of automated driving functions and the safety conception
in an early design stage.
\end{IEEEbiography}

\begin{IEEEbiography}[{\includegraphics[width=1in,clip,keepaspectratio]{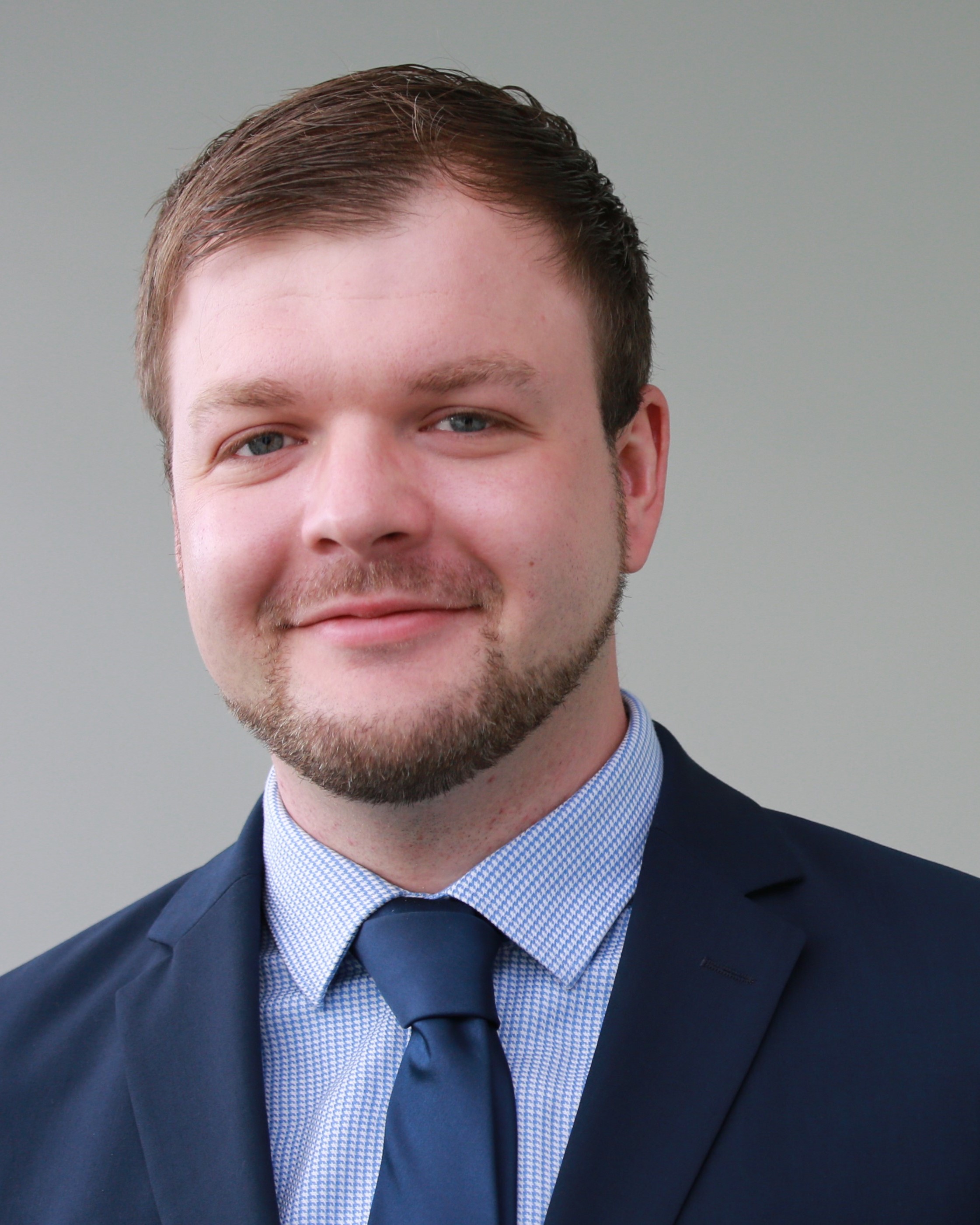}}]{Jan Reich} received his Master's degree in automotive computer science from TU Kaiserslautern, Germany in 2017. Since 2017, he is a full-time researcher in the Safety Engineering department at the Fraunhofer Institute for Experimental Software Engineering (IESE). In his current role as scientific expert Dynamic Assurances of Connected Autonomous Systems, he orchestrates research concerning the development and assurance of intelligent runtime safety monitors using sophisticated environment and self-awareness.
\end{IEEEbiography}

\begin{IEEEbiography}[{\includegraphics[width=1in,clip,keepaspectratio]{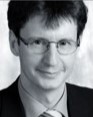}}]{Markus Maurer} received the Diploma degree in electrical engineering from the Technische Universität München, in 1993, and the PhD degree in automated driving from the Group of Prof. E. D. Dickmanns, Universität der Bundeswehr München, in 2000. 
From 1999 to 2007, he was a Project Manager and the Head of the Development Department of Driver Assistance Systems, Audi. 
Since 2007, he has been a Full Professor of automotive electronics systems with the Institute of Control Engineering, Technische Universität Braunschweig.
His research interests include both functional and systemic aspects of automated road vehicles.
\end{IEEEbiography}

\EOD

\end{document}